%% file: OSFTcyclicCoho.tex
\documentclass[11pt]{article}
\usepackage{amssymb,epsf,graphicx}
\usepackage{amsmath}

\allowdisplaybreaks[1]

\newcommand{\nn}{\nonumber}

\newcommand{\eps}{\epsilon}
\newcommand{\de}{\hbox{d}}

\begin{document}

\begin{titlepage}
{}~
\hfill{LMU-ASC 79/10}

\vspace*{2.0cm}

\centerline{\Large \bf Closed String Cohomology in Open String Field Theory}

\vspace*{6.0ex}
\centerline{\Large Nicolas Moeller\footnote{
E-mail: {\tt nicolas.moeller@physik.uni-muenchen.de}} 
and Ivo Sachs\footnote{
E-mail: {\tt Ivo.Sachs@physik.uni-muenchen.de}}
}

\vspace*{0.3cm}

\centerline{\large {\it Arnold Sommerfeld Center for Theoretical Physics,}}
\centerline{\large {\it Theresienstrasse 37, D-80333 Munich, Germany}}

\vspace*{10.0ex}        

\centerline{\bf Abstract}
\bigskip

We show that closed string states in bosonic string field theory are
encoded in the cyclic cohomology of cubic open string field theory
(OSFT) which, in turn, classifies the deformations of OSFT. This
cohomology is then shown to be independent of the open string
background. Exact elements correspond to closed string gauge
transformations, generic boundary deformations of Witten's 3-vertex
and infinitesimal shifts of the open string background. Finally it is
argued that the closed string cohomology and the cyclic cohomology of
OSFT are isomorphic to each other.

\end{titlepage}

\tableofcontents

%%%%%%%%%%%%%%%%%%%%%%%%%%%%%%%%%%%%%%%%%%%%%%%%%%%%%%%%%%%%%%%%%%%%%%
\section{Introduction}
%%%%%%%%%%%%%%%%%%%%%%%%%%%%%%%%%%%%%%%%%%%%%%%%%%%%%%%%%%%%%%%%%%%%%%

It has been known for a long time that Witten's open bosonic string
field theory without the inclusion of closed strings cannot be
unitary. Indeed the perturbative one-loop open string diagrams contain
closed string poles. One strategy is then to extend OSFT by adding
closed and open-closed string vertices explicitly. This results in
Zwiebach's open-closed string field theory
\cite{Zwiebach:1990qj,Zwiebach:1997fe}. Another strategy (although not
necessarily orthogonal to the first one) is to identify the cohomology
of closed string states in cubic open string field theory. To be more
precise we recall that open string field theory is based on a
differential graded algebra (DGA) ${\cal{A}}=(Q,*,A)$, where the odd
differential $Q$ is the open string BRST operator. Its cohomology is
that of the open string states. The idea is then to identify a second
cohomology in OSFT whose elements are the closed string states. Now,
it was shown \cite{Gaberdiel:1997ia} that self-consistent gauge
invariant generalizations of cubic SFT are based on
$A_\infty$-algebras admitting an invariant inner product $\langle a*b
, c \rangle = \langle a , b*c \rangle$. On the other hand it is well
known (see e.g \cite{Penkava:1994mu} ) that infinitesimal
$A_\infty$-deformations of a DGA preserving the invariant inner
product, are classified by cyclic cohomology. This suggests that the
closed string cohomology should be related to the cyclic cohomology of
the differential graded algebra of cubic string field theory (see also
\cite{Kapustin:2004df}). To make this statement precise we then
consider the vertices of Zwiebach's open-closed string field theory
consisting of disks with a single closed string insertion in the bulk
and an arbitrary number of open string insertions on the boundary. We
then show that the condition for these vertices to be elements of the
cyclic cohomology is precisely that the closed string insertion is
on-shell.

\paragraph{}
Let us summarize our results. We consider arbitrary infinitesimal
deformations of Witten's cubic string field theory
$$
S = \frac{1}{2} \, \langle \Psi, Q \Psi \rangle + 
\frac{1}{3} \, \langle \Psi, \Psi, \Psi \rangle + 
\sum_{M=1}^\infty C_M (\Psi, \ldots, \Psi).
$$
We know that they must satisfy $A_\infty$-algebraic conditions or,
equivalently, the multilinear maps $C_M$ must be classified by the
Hochschild cohomology. Let us briefly review what this means. First
the BRST charge $Q$ acts on $C_M$ in the following way
$$
(QC_M)(\Psi_1, \ldots, \Psi_M) \equiv 
\sum_{i=1}^M (-1)^{\Psi_1 + \ldots + \Psi_{i-1}} C_M(\Psi_1,\ldots, \Psi_{i-1},
Q\Psi_i, \Psi_{i+1}, \ldots, \Psi_M).
$$
And the co-boundary operator $\delta: \, \text{Hom}(A^{M-1},{\mathbb
 C})\to \text{Hom}(A^{M},{\mathbb C})$ is defined by
\begin{align}
& (\delta C_{M-1})(\Psi_1, \ldots, \Psi_{M-1}, \Psi_M) \nonumber \\ 
= & \, \sum_{i=1}^{M-1} (-1)^i C_{M-1}(\Psi_1, \ldots, \Psi_i*\Psi_{i+1},\ldots,\Psi_M)
+ (-1)^{\Psi_1(\Psi_2+\ldots +\Psi_M)} C_{M-1}( \Psi_2, \ldots,\Psi_M*\Psi_1) \nonumber.
\end{align}
It turns out that $Q^2=0$, $\delta^2=0$ and $[Q,\delta]=0$. And
therefore the operator $(\delta - (-1)^M Q)$ squares to zero and thus
defines a complex. The Hochschild cohomology is the cohomology of
$(\delta - (-1)^M Q)$. This is almost what we need, but not quite
exactly; we will need to focus on the "cyclic" elements of the
cohomology. Namely we must ask that the open string maps $C_M$ satisfy
$$
C_M(\Psi_2, \ldots, \Psi_M, \Psi_1) = (-1)^{M-1+\Psi_1(\Psi_2 + \ldots + \Psi_M)} 
C_M(\Psi_1, \ldots, \Psi_M),
$$
and the Hochschild cohomology restricted to the cyclic maps is called
the cyclic cohomology ($HC^M$).

The main result of this paper is that the closed string states can be
identified with this cohomology. The proof goes as follows. First,
remember that there is an isomorphism between cyclic cohomology and
infinitesimal deformations of cubic SFT. Next, we list the possible
deformations; we find three possibilities:
\begin{enumerate}
\item Deformations of the closed string background
\item Deformations of the open string background
\item Generic deformations of the three-point vertex
\end{enumerate}
We will find that to each physical closed string state, there
corresponds a closed element of the cyclic complex $CC^M$. Moreover,
two closed string states are gauge-equivalent if and only if their
elements in the cyclic complex are equivalent. In other words, there
is a one-to-one linear map from the space of physical closed string
states to the cyclic cohomology $HC^M$. In contrast, we will show that
deformations of the open string background correspond to trivial
elements in $HC^M$. The generic deformations of the 3-vertex are a
little more subtle, but we will show that the geometric deformations
are exact. And we will argue that all other deformations of the
3-vertex are either trivial or correspond to a closed string state of
ghost number zero. Since the semi-relative closed string cohomology at
ghost number zero and vanishing mass$^2$ is trivial we can then
conclude that all elements in the cyclic cohomology must correspond to
physical closed string states, and the main result of the paper
follows.

\paragraph{}
We then consider finite deformations of the open string
background. Namely we expand the theory around a solution of the
classical equations of motion and look at the cyclic cohomology
there. We are able to prove that the cyclic cohomology around any
finite background is isomorphic to the original one.

\paragraph{}
What is the relevance of our results? On a conceptual level they show
that OSFT already encodes the closed string states. Indeed the
cohomology introduced here is defined using only the structure of
OSFT, i.e. the open string BRST operator $Q$ and the product of open
string fields. No extra information about closed strings enter
here. On the other hand, we will argue below that Zwiebach's
open-closed vertices are the only possible non-trivial generalizations
of Witten's 3-vertex (assuming local insertions on the boundary). In
that sense the open-closed string vertices are the only non-trivial
deformations of OSFT. We will return to this point again in the
conclusions. Finally, because of the fact that cyclic 
cohomology is invariant under deformations of the open string
background, it classifies the open-closed string vertices in any open
string background, in particular in the tachyon vacuum where the open
string cohomology is trivial.

\paragraph{}
Plan of this paper: In section \ref{cocsft} we recall the relevant
aspects of the construction of open-closed string field theory.  In
section \ref{sft-cyclic} we show that the closed string cohomology is
contained in the cyclic cohomology of the differential graded algebra
${\cal{A}}$, while closed string gauge transformation are exact
elements in the cyclic complex. In section \ref{un} we then argue that
the closed string cohomology and the cyclic cohomology of ${\cal{A}}$
are in fact isomorphic to each other. In section \ref{bi} we show that
the cyclic cohomology of ${\cal{A}}$ is independent of the open string
background. In section \ref{conc} we present the conclusions and some
open issues. In appendix \ref{HH} we review some facts about
$A_\infty$-algebras drawing on an analogy with algebraic topology.

%%%%%%%%%%%%%%%%%%%%%%%%%%%%%%%%%%%%%%%%%%%%%%%%%%%%%%%%%%%%%%%%%%%%%%
\section{Linearized Open-Closed SFT}
\label{cocsft}
%%%%%%%%%%%%%%%%%%%%%%%%%%%%%%%%%%%%%%%%%%%%%%%%%%%%%%%%%%%%%%%%%%%%%%

Zwiebach's quantum open-closed string field theory with manifest
closed string factorization \cite{Zwiebach:1990qj, Zwiebach:1997fe} is
given by the action
\begin{equation}
S = \sum_{p=0,\frac{1}{2},1,\ldots} \hbar^p S_p,
\end{equation}
with
\begin{equation}
S_p = \sum_{\genfrac{}{}{0pt}{}{G,N,B}{2G+B-1+N/2=p}} 
\sum_{\genfrac{}{}{0pt}{}{m_1,\ldots,m_B \geq 0}{N+M \geq 1}} g^{2p+N-2+M} S_{B,M}^{G,N}.
\end{equation}
The $S_{B,M}^{G,N}$ are contact terms described by genus $G$ surfaces
with $N$ closed string insertions in the bulk, $B$ boundaries and
$m_k$ open string insertions on the $k$-th boundary, and with $M = m_1
+ \ldots + m_B$. All the vertices in this action include strips of
length $\pi$ for the external open strings, and stubs of length $\pi$
for the external closed strings. This is necessary at the quantum
level to avoid over-counting of moduli space as, for example, an open
string loop diagram with a short internal propagator is
equivalent to a diagram with a long internal closed string
propagator. The strips and stubs avoid the appearance of propagators
of length smaller than $2\pi$. But this implies that Witten's vertex
is modified, and contact vertices with $M$ open strings on the
boundary of the disk must then be included for all $M \geq 4$.

We will be interested in the classical limit of the action $\hbar
\rightarrow 0$. The dominant term $S_0$ contains the free part of the
closed SFT action (which will not be relevant to us), and all purely
open string vertices with strips. We will also consider the
next-to-leading term $S_\frac{1}{2}$. It consists of the purely closed
cubic vertex (which again will not be relevant), and all the contact
terms described by a disk with one closed string insertion and $M$
open string insertions on the boundary, where
$M=0,1,\ldots,\infty$. And we will neglect all other vertices because
they are of order one or higher in $\hbar$. To this order, it is
consistent to consider the closed strings as non-dynamical; in
particular the closed string tadpole ($M=0$) is just a constant that
we can ignore.

In this limit, it is consistent to replace the purely open string
sector by the Witten vertex without external strips. To see why, we
note that the over-counting mentioned above concerns diagrams with at
least one internal closed string propagator. These diagrams can arise
in two different ways. First as Feynman diagrams built from at least
two vertices with one closed strings and open strings on the
boundary. These are of order $\hbar^\frac{1}{2}$ or higher and
therefore the diagram will be of order at least one in
$\hbar$. The other possibility is that the diagram is a contact term
with two or more boundaries and therefore also of order at least one
in $\hbar$. In the classical limit, we neglect these diagrams and we
can thus consistently describe the purely open sector by Witten's
cubic action without strips.

\begin{figure}
\begin{center}
\input{ocdisk.pstex_t}
\caption{The correlator $C_M^{\phi}(\Psi_1,\ldots,\Psi_M)$ on the disk
 $D$ with global coordinate $w$. The open string punctures are at the
 points $w_i = e^{is_i}$.}
\label{ocdisk}
\end{center}
\end{figure}
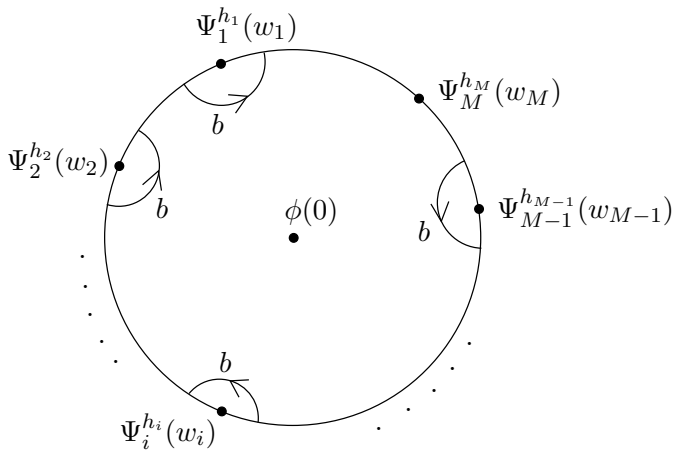

We will then focus on the contact terms $S^{0,1}_{1,M}$ coupling one
closed string and M open strings on the disk. It is represented by the
integrated correlator $C^\phi_M(\Psi_1, \Psi_2, \ldots, \Psi_M)$
defined as in Fig.\ref{ocdisk}.
We have a disk with one puncture in the middle and M punctures on the
boundary, and local coordinates around each punctures defined by the
quadratic differential solving the minimal area problem with the
constraints that all non-trivial open curves have length $\geq \pi$
and all non-trivial closed curves have length $\geq 2 \pi$.  
The vertex operators $\phi$ and ${\Psi_i}$ are inserted in the local
coordinates. Since the moduli space of our punctured disk has
dimension $M-1$, we need to insert $M-1$ antighosts that will
determine the measure on the moduli space. We do that by inserting
line integrals of $b$ around each but one of the open string
vertex operators. Using the doubling trick, we can write them as
\begin{equation}
b(v_i) = \frac{1}{2 \pi i} \oint_{P_i} b(z) \, v_i(z) \, dz.
\end{equation} 
Let's comment on that:  
The vector $v(z)$ must correspond to a tangent vector in the moduli
space. This is the Schiffer variation argument that we briefly
sketch. In our case, the disk with one puncture in the center and $M$
punctures on the boundary, has $M-1$ moduli which can be taken as the
positions of $M-1$ punctures on the boundary. We know then that
there exist $M-1$ meromorphic vector fields $v_i$ defined in a
neighborhood of the puncture $P_i$ which are constant near $P_i$, and
which cannot be extended to the whole disk. It turns out that these
vector fields generate translations in the moduli space; in other
words, in our case they move the punctures. From the operator
formalism point of view, these translations are generated by the line
integrations of the energy-momentum tensor weighted by the vector
$v_i$
\begin{equation}
T(v_i) \equiv \frac{1}{2 \pi i} \oint_{P_i} T(z) \, v_i(z) \, dz.
\end{equation}
We can now write the definition of the integrated correlator:
\begin{equation}
C^\phi_M(\Psi_1, \Psi_2, \ldots, \Psi_M) \equiv 
\int_{{\cal T}_M} ds_1 \ldots ds_{M-1} \, \langle \phi(0) \, b(v_1) 
\ldots b(v_{M-1}) \, \Psi^{h_1}_1(w_1) \ldots \Psi^{h_M}_M(w_M) \rangle_D.
\label{Cdef}
\end{equation}
Here ${\cal T}_M$ is the part of moduli space of the disk with one
puncture in the bulk and $M$ punctures on the boundary, not covered by
Feynman diagrams obtained by lower order vertices. The $s_i$ are
$M-1$ real numbers parameterizing the moduli space. 
Concretely, we will take the coordinate on the disk to be 
\begin{equation}
w = re^{i s},
\end{equation}
so the $s_i$ are the angles specifying the positions of the open
string punctures, namely $w_i = e^{i s_i}$. Finally,
$\Psi^{h_i}_i(w_i) \equiv h_i \circ \Psi(0)$ denotes the vertex
operator inserted in the local coordinates with the conformal map
$h_i$. The only thing that we will need to know about $h_i$ is that it
depends on all $s_k$ ($k=1,\ldots,M$) and that $h_i(0) = w_i = e^{i
  s_i}$.

%%%%%%%%%%%%%%%%%%%%%%%%%%%%%%%%%%%%%%%%%%%%%%%%%%%%%%%%%%%%%%%%%%%%%%
\section{Closed Strings and Cyclic Cohomology}\label{sft-cyclic}
%%%%%%%%%%%%%%%%%%%%%%%%%%%%%%%%%%%%%%%%%%%%%%%%%%%%%%%%%%%%%%%%%%%%%%

Let us now consider Witten's open string field theory (in a flat
closed string background) and let us switch on an infinitesimal closed
string background $\phi$. The theory is now described by the OCSFT
action
\begin{equation}
S = \frac{1}{2} \, \langle \Psi, Q \Psi \rangle + 
\frac{1}{3} \, \langle \Psi, \Psi, \Psi \rangle + 
\sum_{M=1}^\infty C^{\phi}_M (\Psi, \ldots, \Psi)
\end{equation}
which we now view as an infinitesimal deformation of the cubic OSFT.
The closed string background $\phi$ being infinitesimal, its equation
of motion is simply
\begin{equation}
Q |\phi\rangle = 0.
\label{Qphi}
\end{equation}
On the other hand, every on-shell physical closed string state $\phi$
obeys Eq.~(\ref{Qphi}), and can therefore be used to produce an
infinitesimal background deformation. However, Eq.~(\ref{Qphi}) is not
the only condition that $\phi$ must fulfill in order to be a physical
state. It is well known that the condition
\begin{equation}
b_0^- |\phi \rangle = 0
\label{b0}
\end{equation}
must be imposed as well, where $b_0^- \equiv b_0 - \bar{b}_0$. Then
the last condition
\begin{equation}
L_0^- |\phi \rangle = 0,
\end{equation}
where $L_0^- \equiv L_0 - \bar{L}_0$, follows immediately after acting
with $Q$ on Eq.~(\ref{b0}).

We will now show that the constraint (\ref{b0}) implies that the
multilinear functions $C^{\phi}_M$ have the cyclicity symmetry
\begin{equation}
C^\phi_M( \Psi_2, \ldots, \Psi_M,\Psi_1) =(-1)^{M-1+\Psi_1(\Psi_2+\ldots 
+\Psi_M)}C^{\phi}_M( \Psi_1, \ldots, \Psi_{M-1},\Psi_M).
\label{cycl}
\end{equation}
To show this, we note that the vectors $v_i$ in the definition of the
correlator, are given by
\begin{equation}
v_i = -\partial_{s_i}
\end{equation}
in the local coordinate patch at $s_i$, and $v_i = 0$ around the other
punctures.\footnote{The minus sign comes form the Schiffer variation
 argument.} The sum of these $M$ vector fields can be extended to the
whole disk, it is simply given by
\begin{equation}
v = \sum_{i=1}^M v_i = -\partial_s.
\end{equation}
And remembering that the coordinate on the disk is $w = re^{is}$, we have that 
\begin{equation}
\partial_s = i ( w \partial_w - \bar{w} \partial_{\bar{w}}).
\end{equation}
It follows that 
\begin{equation}
\sum_{i=1}^M b(v_i) = - \oint_0 b(w)v(w)\frac{dw}{2\pi i} 
- \oint_0 \bar{b}(\bar{w})\bar{v}(\bar{w})\frac{d\bar{w}}{2\pi i} 
= i \, b_0^-,
\label{bb0}
\end{equation}
where it is understood that $b_0^-$ acts at $w=0$, i.e. on the closed
string field $\phi$. Let us now consider the integrated correlator
with $\phi$ replaced by $b_0^-\phi$, and with one less line integral
of $b$ (for concreteness we remove $b(v_{M-1})$), and also we
cyclically permute $\Psi_1$ to the right of $\Psi_M$:
\begin{equation}
\int_{{\cal T}_M} ds_1 \ldots ds_{M-1} \, \langle (b_0^- \phi)(0) 
\, b(v_1) \ldots b(v_{M-2}) \, \Psi^{h_1}_2(w_1) 
\ldots \Psi^{h_{M-1}}_M(w_{M-1}) \Psi^{h_M}_1(w_M) \rangle_D.
\label{b0psi}
\end{equation}
Using (\ref{bb0}), we see that this is proportional to
\begin{align}
& \int_{{\cal T}_M} ds_1 \ldots ds_{M-1} \, \langle \phi(0) \, b(v_1) 
\ldots b(v_{M-2}) \, \sum_{l=1}^M b(v_l) \, \Psi^{h_1}_2(w_1) 
\ldots \Psi^{h_{M-1}}_M(w_{M-1}) \Psi^{h_M}_1(w_M) \rangle_D \nonumber\\
& = \int_{{\cal T}_M} ds_1 \ldots ds_{M-1} \, \langle \phi(0) \, b(v_1) 
\ldots b(v_{M-2}) b(v_{M-1}) \, \Psi^{h_1}_2(w_1) 
\ldots \Psi^{h_{M-1}}_M(w_{M-1}) \Psi^{h_M}_1(w_M) \rangle_D \nonumber\\
& \quad + \int_{{\cal T}_M} ds_1 \ldots ds_{M-1} \, \langle \phi(0) \, b(v_1) 
\ldots b(v_{M-2}) b(v_M) \, \Psi^{h_1}_2(w_1) 
\ldots \Psi^{h_{M-1}}_M(w_{M-1}) \Psi^{h_M}_1(w_M) \rangle_D \nonumber\\
& = C_M^{\phi}(\Psi_2, \ldots,\Psi_M, \Psi_1) + (-1)^{M-2 + \Psi_1 (\Psi_2 + 
\ldots +\Psi_M)} \times \nonumber \\
& \quad \times \int_{{\cal T}_M} ds_1 \ldots ds_{M-1} \, \langle \phi(0) \, 
b(v_M) b(v_1) \ldots b(v_{M-2}) \, \Psi^{h_M}_1(w_M) \Psi^{h_1}_2(w_1) 
\ldots \Psi^{h_{M-1}}_M(w_{M-1}) \rangle_D. \label{Ccyclic}
\end{align}
In the first step we have replaced $i b_0^-$ acting on $\phi$, by
$\sum_{l=1}^M b(v_l)$ acting on the open strings. And in the last
line, we have moved both $b(v_M)$ and $\Psi^{h_M}_1(w_M)$ to the left
and accounted for the corresponding sign.  In order to identify the
last term, we note that the domain of integration ${\cal T}_M$ is
invariant upon cyclic permutation. Indeed, until now we have chosen to
keep $w_M$ fixed to a given (unspecified) value, but the correlator is
invariant under a global rotation of the disk. This means that we
could equivalently have chosen to keep, say, $w_{M-1}$ fixed; we would
then have to integrate over $ds_1 \ldots ds_{M-2} ds_M$. With this
remark, we see that (\ref{Ccyclic}) is
$$
C_M^{\phi}(\Psi_2, \ldots,\Psi_M, \Psi_1) +
(-1)^{M-2+\Psi_1(\Psi_2+\ldots +\Psi_M)}C^\phi_M( \Psi_1, \ldots, \Psi_{M-1},\Psi_M).
$$
Since we have kept the $\Psi_i$'s completely arbitrary, (\ref{b0psi})
is identically zero if and only if $b_0^- \, |\phi \rangle = 0$, and
we can conclude that
\begin{align}
& b_0^- \, |\phi \rangle = 0 \qquad \text{if and only if} \nonumber \\
& C^\phi_M( \Psi_2, \ldots, \Psi_M,\Psi_1) =(-1)^{M-1+\Psi_1(\Psi_2+
\ldots +\Psi_M)}C^\phi_M( \Psi_1, \ldots, \Psi_{M-1},\Psi_M).
\end{align}

\paragraph{}
We are now ready to look at the equation of motion and gauge symmetry
of the closed string sector, and to try to relate them to algebraic
conditions on the open string vertices.

\begin{figure}
\begin{center}
\input{overlap.pstex_t}
\caption{The overlap pattern}
\label{overlap}
\end{center}
\end{figure}
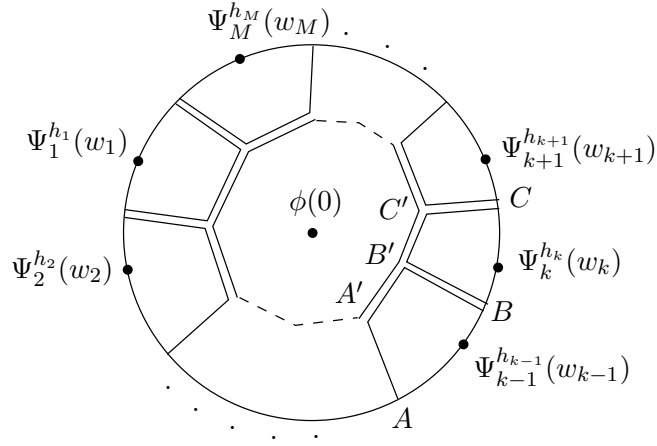

Let us act with the BRST charge $Q$ on the closed string field
$\phi$. Note that we don't put any restriction on the ghost numbers of
the string fields, so that inserting $Q$ in the vertex doesn't give
zero by trivial ghost number counting. This can be represented on the
disk by inserting an integration of the BRST current $j$ along a
closed contour around the origin (the closed string puncture). We can
further deform the contour by pushing it towards the boundary of the
disk and then splitting it to $M$ contours around the open string
punctures and $b(v_i)$ insertions. We can cross the $b(v_i)$
insertions, thereby producing a $T(v_i)$ by virtue of
\begin{equation}
\left\{ Q, b(v_i) \right\} = T(v_i),
\label{QbL}
\end{equation}
As already mentioned, $T(v_i)$ generates translations in the moduli
space; more precisely:
\begin{equation}
\langle T(v_i) \ldots \rangle_D = - \frac{\partial}{\partial s_i} \langle \ldots \rangle_D.
\label{Taction}
\end{equation}
All in all, using the definition (\ref{Cdef}),
we have
\begin{align}\label{qfc}
& \quad C^{Q \phi}_M( \Psi_1, \ldots, \Psi_{M-1}, \Psi_M) \nn\\
& = \int_{{\cal T}_M} ds_1 \ldots ds_{M-1}
\langle Q \, \phi(0) b(v_1)\ldots b(v_{M-1})\, \Psi^{h_1}_1(w_1) \ldots 
\Psi^{h_{M-1}}_{M-1}(w_{M-1}) \Psi^{h_M}_M(w_M) \rangle_D \nn\\
& = (-1)^{1+\phi} \int_{{\cal T}_M} ds_1 \ldots ds_{M-1}
\langle \phi(0) \, Q \, b(v_1) \ldots b(v_{M-1}) \, \Psi^{h_1}_1(w_1) \ldots 
\Psi^{h_{M-1}}_{M-1}(w_{M-1}) \Psi^{h_M}_M(w_M)\rangle_D \nn\\
& = (-1)^{1+\phi} \int_{{\cal T}_M} ds_1 \ldots ds_{M-1} \,
\sum_{l=1}^{M-1}(-1)^{l} \times \nn \\
& \qquad \qquad \qquad \times \frac{\partial}{\partial s_l}
\langle \phi(0) \, b(v_1) \ldots b(v_l) \!\!\!\!\!\!\!\!\!\!\!\!/ \,\;\;\;\;\; 
\ldots b(v_{M-1}) \, 
\Psi^{h_1}_1(w_1) \ldots \Psi^{h_l}_l(w_l) \ldots \Psi^{h_M}_M(w_M) \rangle_D \nn\\
& \quad + (-1)^{1+\phi} \, \sum_{l=1}^M(-1)^{M-1+\Psi_1+\ldots +\Psi_{l-1}} 
C^{\phi}_M( \Psi_1, \ldots, Q \Psi_{l},\ldots,\Psi_M).
\end{align}
In the second and third lines, we have replaced the closed BRST
operator $Q$, acting on $\phi$, with the open BRST operator acting on
the boundary. In doing so, we must account for a minus sign coming
from the change of orientation of the contour of integration defining
$Q$, and also for a $(-1)^\phi$ coming from moving $Q$ to the right of
$\phi$.\footnote{Equivalently, note that $\text{bpz}(Q|\phi\rangle) =
  (-1)^{1+\phi} \langle \phi| Q$.}  In the fourth line, we are
integrating total derivatives, and we will thus have contributions
only from the boundary of moduli space $\partial {\cal T}_M$. Let us
be a little more explicit about these last contributions. The domain
of integration of $s_k$, when all other $s_i$ are fixed, is some
interval $(s_k^<,s_k^>)$ strictly contained in the interval
$(s_{k-1},s_{k+1})$, When $s_k$ is at a boundary of the interval, e.g.
$s_k=s_k^<$, then the punctured disk is identical to the Riemann
surface given by the Feynman diagram obtained by gluing the Witten
cubic vertex to the disk with $M-1$ punctures via a propagator of
length zero. Let us look at this from the point of view of overlap
patterns. In Figure~\ref{overlap}, we show the decomposition of the
disk into one semi-infinite cylinder and $M$ semi-infinite strips of
width $\pi$, arising from a Jenkins-Strebel quadratic differential
\cite{Zwiebach:1997fe}. The condition for this Riemann surface to
belong to the moduli space, is that all non-trivial closed curves are
not shorter than $2 \pi$ and that all non-trivial open curves are not
shorter than $\pi$. In particular the segment $AA'B'C'C$ must have
length of at least $\pi$; this implies that the length of the segment
$BB'$, which we call $a_{k-1,k}$ must satisfy $a_{k-1,k} \leq
\frac{\pi}{2}$. When $s_k=s_k^<$, we are at a boundary of the moduli
space characterized by $a_{k-1,k} = \frac{\pi}{2}$; this means that
the strings $\Psi_{k-1}$ and $\Psi_k$ overlap exactly on their right
and left halves respectively.\footnote{As explained in
  \cite{Zwiebach:1997fe}, $a_{k-1,k}=0$ is not a boundary of the
  moduli space.} This is precisely the star product. In other words,
the correlator is unchanged if we remove the strips corresponding to
$\Psi_{k-1}$ and $\Psi_k$, and replace them by a strip corresponding
to $\Psi_{k-1} * \Psi_k$.

We thus end up with
\begin{align}\label{third}
&\int_{{\cal T}_M} d^{M-1}s \,
\sum\limits_{l=1}^{M-1}(-1)^{l} \frac{\partial}{\partial s_l}
\langle \phi(0) b(v_1) \ldots b(v_l) \!\!\!\!\!\!\!\!\!\!\!\!/\,\,\,\;\;\; 
\ldots b(v_{M-1}) \, 
\Psi^{h_1}_1(w_1) \ldots \Psi^{h_l}_l(w_l) \ldots \Psi^{h_M}_M(w_M) \rangle_D \nn\\
&= C^{ \phi}_{M-1}( *\Psi_1, \Psi_2,\ldots, \Psi_M)+  \sum
\limits_{l=1}^{M-1}(-1)^{l}C^{ \phi}_{M-1}( \Psi_1, \ldots, \Psi_{l}*\Psi_{l+1},\ldots,\Psi_M)\\
&=(-1)^{\Psi_1(\Psi_2+\ldots +\Psi_M)}C^{\phi}_{M-1}( \Psi_2, \ldots,\Psi_M*\Psi_1)+  
\sum\limits_{l=1}^{M-1}(-1)^{l}C^{\phi}_{M-1} 
(\Psi_1, \ldots, \Psi_{l}*\Psi_{l+1},\ldots,\Psi_M)\nn.
\end{align}
At first sight, it may seem that each product $\Psi_{l}*\Psi_{l+1}$
arises twice with the same sign, once when $s_l = s_l^>$ and once when
$s_{l+1} = s_{l+1}^<$. However, the first contribution comes with a
$b$ integration around $\Psi_{l+1}$ and no $b$ integration around
$\Psi_{l}$, while the situation is opposite for the second
contribution. Therefore, they add up to one term with a $b$
integration path around both $\Psi_{l+1}$ and $\Psi_l$, or in other
words around $\Psi_{l}*\Psi_{l+1}$.  The right hand side of
Eq.~(\ref{third}) is recognized as the Hochschild differential (see
appendix \ref{hccw})
\begin{equation}
\delta: \quad \text{Hom}(A^{M-1},{\mathbb R})\to \text{Hom}(A^{M},{\mathbb R})
\end{equation}
for ${\mathbb Z}_2$ graded algebras. Adding the last term in (\ref{qfc}) we then end up with
\begin{equation}
C^{ Q\phi}_{M}( \Psi_1, \ldots,\Psi_M)= (-1)^{1+\phi} \left(
(\delta C^{ \phi}_{M-1})( \Psi_1, \ldots, \Psi_M) - 
(-1)^M (Q C^{ \phi}_{M})( \Psi_1, \ldots,\Psi_M) \right)
\end{equation}
where $QC_M^\phi$ is defined by
\begin{equation}
(QC_M^{\phi})(\Psi_1, \ldots, \Psi_M) \equiv 
\sum_{i=1}^M (-1)^{\Psi_1 + \ldots + \Psi_{i-1}} C_M^{\phi}(\Psi_1,\ldots, \Psi_{i-1},
Q\Psi_i, \Psi_{i+1}, \ldots, \Psi_M).
\end{equation}
If we impose the condition that the closed string field be on-shell,
$Q \phi = 0$, we then conclude that the maps $(C_1^{\phi},
C_2^{\phi},\ldots, C_M^\phi, \ldots)$ are closed with respect to
$(\delta-(-1)^M Q)$. If the closed string field is off-shell, $Q \phi
\neq 0$ then it is not hard to show that $ C^{ Q\phi}_{M}( \Psi_1,
\ldots,\Psi_M)$ is cyclic, so that the operator $(\delta-(-1)^M Q)$
takes cyclic elements into cyclic elements. Moreover, the above
calculation also shows that if $\phi$ is pure gauge, i.e.  $\phi = Q
\Lambda$ for some $\Lambda$ satisfying $b_0^- \Lambda =0$ and $L_0^-
\Lambda = 0$, then 
\begin{equation}
C^{\phi}_{M}( \Psi_1, \ldots,\Psi_M)= (-1)^{\phi} \left(
(\delta C^{\Lambda}_{M-1})( \Psi_1, \ldots, \Psi_M) - 
(-1)^M (Q C^{\Lambda}_{M})( \Psi_1, \ldots,\Psi_M) \right).
\end{equation}
In other words, it is exact with respect to $(\delta-(-1)^M
Q)$. Furthermore, $(\delta-(-1)^M Q)^2 = 0$. We conclude that the
on-shell closed string states are contained in the cyclic
cohomology\footnote{See appendix \ref{HH} for a definition.} of
$(\delta-(-1)^M Q)$.

%%%%%%%%%%%%%%%%%%%%%%%%%%%%%%%%%%%%%%%%%%%%%%%%%%%%%%%%%%%%%%%%%%%%%%
\section{Generic Deformations and the Cyclic Complex}\label{un}
%%%%%%%%%%%%%%%%%%%%%%%%%%%%%%%%%%%%%%%%%%%%%%%%%%%%%%%%%%%%%%%%%%%%%%

We have seen above that on-shell closed string insertions correspond
to elements in the cyclic cohomology, whereas infinitesimal closed
string gauge transformations correspond to exact elements in the
cyclic complex. In this section we want to argue that this
correspondence is in fact an isomorphism.  To see this we consider
below generic deformations of OSFT (or, equivalently, its
corresponding DGA) that do not correspond to a closed string insertion
in the bulk.

%%%%%%%%%%%%%%%%%%%%%%%%%%%%%%%%%%%%%%%%%%%%%%%%%%%%%%%%%%%%%%%%%%%%%%
\subsection{Deformation of the Open String Background}
%%%%%%%%%%%%%%%%%%%%%%%%%%%%%%%%%%%%%%%%%%%%%%%%%%%%%%%%%%%%%%%%%%%%%%

In this subsection we consider the first class of such deformations
corresponding to an insertion of an open string background on the
boundary of the disc with the open string being on-shell.

\paragraph{}
Let $\Psi_0$ be an infinitesimal marginal open string deformation. We
express it as $\Psi_0 = O(\eps)$. And $\Psi_0$ is a solution of the
equations of motion, therefore $Q \Psi_0 = 0 + O(\eps^2)$. Writing
$\Psi = \Psi_0 + \tilde{\Psi}$, the action becomes
\begin{equation}
S' = \frac{1}{2} \langle \tilde{\Psi}, Q' \tilde{\Psi} \rangle + \frac{1}{3} 
\langle \tilde{\Psi}, \tilde{\Psi}, \tilde{\Psi} 
\rangle + O(\eps^2),
\end{equation}
where the new BRST operator $Q'$ is given by
\begin{equation}
Q'\tilde{\Psi} = Q \tilde{\Psi} + \Psi_0 * \tilde{\Psi} + (-1)^{1 + \tilde{\Psi}} 
\tilde{\Psi} * \Psi_0.
\end{equation}
And from now on we will write $\Psi$ instead of $\tilde{\Psi}$. Next, we rewrite 
the action $S'$ as $S$ plus an infinitesimal deformation
\begin{equation}
S' = S + C_2(\Psi,\Psi) + O(\eps^2),
\label{SC2}
\end{equation}
where $C_2$ is defined by
\begin{equation}
C_2(\Psi_1, \Psi_2) = \frac{1}{2} \left( \langle \Psi_0, \Psi_1, \Psi_2 
\rangle + 
(-1)^{1+\Psi_1\Psi_2} \langle \Psi_0, \Psi_2, \Psi_1 \rangle \right).
\label{C2def}
\end{equation}
This definition makes sense because when $\Psi_1 = \Psi_2 = \Psi$,
with $|\Psi|=1$, we indeed have (\ref{SC2}). To verify this fact, one
only needs the following elementary identities
\begin{align}
& \langle A, B, C \rangle = (-1)^{A(B+C)} \langle B, C, A \rangle \\
& \langle A, B, C \rangle = \langle A*B, C \rangle = \langle A, B*C \rangle.
\end{align}
And we mention here one more identity that will be needed shortly
\begin{equation}
\langle A, B \rangle = (-1)^{AB} \langle B, A \rangle.
\end{equation}

Let us now look at the algebraic properties of $C_2$. From the
definition (\ref{C2def}), we immediately see that $C_2$ satisfies the
cyclicity condition
\begin{equation}
C_2(\Psi_2,\Psi_1) =
(-1)^{1+\Psi_1\Psi_2}C_2(\Psi_1,\Psi_2).
\end{equation}
Furthermore, it is easy to see that the collection of maps formed by
$C_2$ alone, namely $(0, C_2, 0, 0, \ldots)$, is exact.
Indeed, let us define
\begin{equation}
D_1(\Psi) = -\frac{1}{2} \langle \Psi_0, \Psi \rangle.
\end{equation} 
We then have
\begin{equation}
(Q D_1)(\Psi) = -\frac{1}{2} \langle \Psi_0, Q \Psi \rangle = 
-\frac{1}{2} \langle
Q \Psi_0, \Psi \rangle = 0 + O(\eps^2).
\end{equation}
And, straight from the definition of $\delta$, we find that
\begin{equation}
C_2 = \delta D_1.
\end{equation}
Thus we conclude that, to order $\eps^1$, $(0,C_2,0,0,\ldots)$ is
exact, namely
\begin{equation}
(0,C_2,0,\ldots) = (\delta-(-1)^NQ)\, (D_1,0,0, \ldots).
\end{equation}

\paragraph{}
We are thus left with the class of deformations of OSFT obtained by
acting with some operator ${\cal{ O}}$ on the open string fields. We
will treat this case in the next subsection.

%%%%%%%%%%%%%%%%%%%%%%%%%%%%%%%%%%%%%%%%%%%%%%%%%%%%%%%%%%%%%%%%%%%%%%
\subsection{Generic Deformations of the 3-Vertex} \label{GD}
%%%%%%%%%%%%%%%%%%%%%%%%%%%%%%%%%%%%%%%%%%%%%%%%%%%%%%%%%%%%%%%%%%%%%%

Let us assume that $C_n$ is non-vanishing for some $n>3$. Then
$C_n\neq 0$ for some $n\leq 3$ as well for, if $C_n= 0$ for $n\leq 3$
then the moduli space of perturbative OSFT is covered exactly once if
and only if $C_n=0, n>3$.  This is just the statement of consistency
of Witten's OSFT.

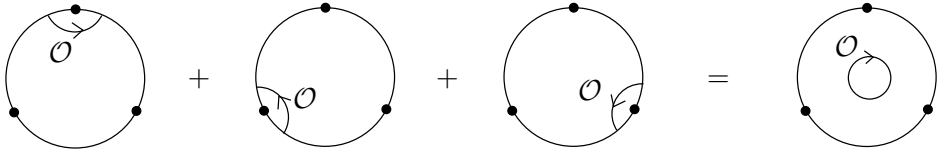
\begin{figure}
\begin{center}
\input{OB.pstex_t}
\caption{ If ${\cal{O}}$ is a conformal invariant this is equivalent
  to a closed contour around the origin.}
\label{OB}
\end{center}
\end{figure}

Let us now focus on possible deformations of the $3$-vertex obtained
by acting with some operator ${\cal{ O}}$ on the open string fields.
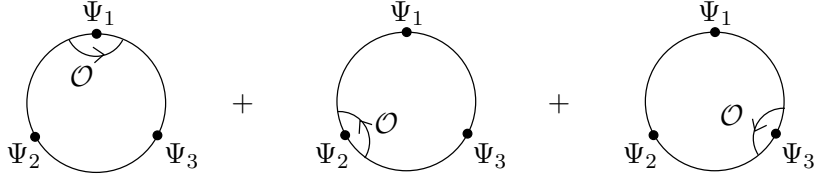
\begin{figure}
\begin{center}
\input{3wd.pstex_t}
\caption{Generic deformation of Witten's 3-vertex by an open string
  operator ${\cal{O}}$ written as a contour integral around the open
  string puncture.}
\label{3wd}
\end{center}
\end{figure}
Cyclicity then requires that we sum over the three graphs on the left
hand side in Fig.~\ref{3wd}. A generic deformation $C_3$ in the cyclic
cohomology is then given by
\begin{equation}
C_3(\Psi_1,\Psi_2,\Psi_3)=\langle {\cal O}\Psi_1,\Psi_2*\Psi_3\rangle
+(-1)^{{\cal O}\Psi_1} \langle\Psi_1, {\cal O}\Psi_2*\Psi_3\rangle 
+ (-1)^{{\cal O}(\Psi_1+\Psi_2)} \langle\Psi_1, \Psi_2*{\cal O}\Psi_3\rangle
\end{equation}
subject to the equation
\begin{equation}\label{cc2}
\delta C_2+QC_3=0.
\end{equation}
Without restricting the generality we can assume that bpz$( {\cal
 O})=\pm {\cal O}$. It is then not hard to see that for ${\cal
 B}=[{\cal O},Q\}\neq 0$ Eq.~(\ref{cc2}) has a solution only if
${\cal O}$ is BPZ-odd and then the solution is given by
\begin{equation}
C_2(\Psi_1,\Psi_2)=-\langle {\cal B}\Psi_1, \Psi_2\rangle
\end{equation}
Note that the BPZ-parity of ${\cal O}$ is the same as that of $ {\cal
 B}$. However, for ${\cal O}$ BPZ-odd, $C_2$ and $C_3$ are exact,
$C_2=-(QD_2)$ and $C_3=\delta D_2$ with $D_2= \langle {\cal
 O}\Psi_1, \Psi_2\rangle$. The remaining possibility is then ${\cal
 B}=0$ and bpz$( {\cal O})= {\cal O}$. Then $QC_3=0$ and we are
left with the condition
\begin{equation}\label{cc3}
\delta C_3+QC_4=0
\end{equation}
with 
\begin{equation}\label{dc3}
\delta C_3=-2\langle{\cal O}(\Psi_1*\Psi_2),\Psi_3*\Psi_4\rangle
+2(-1)^{{\cal O}\Psi_1}\langle\Psi_4*\Psi_1,{\cal O}(\Psi_2*\Psi_3)\rangle
\end{equation}
If ${\cal O}$ is $Q$-exact then $C_3$ is again trivial in the cyclic
cohomology. On the other hand if ${\cal O}$ is in the cohomology of
$Q$, we claim that (\ref{cc3}) can have a solution only if ${\cal O}$
is a conformal invariant so that ${\cal O}$ can be pulled in the bulk
(see Fig.~\ref{OB}). 
Indeed since ${\cal O}$ is not $Q$-exact, the only way the
differential $Q$ acting on $C_4$ can reproduce (\ref{dc3}) is as a
derivative on its moduli space. However, while the boundary of this
moduli space may contain terms of the form $\Psi* {\cal{O}}\Psi$, it
does not contain terms of the form appearing in (\ref{dc3}) unless
${\cal O}$ is a derivative of $*$. Since ${\cal O}$ is BPZ-even it
cannot be a derivative. On the other hand if ${\cal O}$ can be pulled
in the bulk as in Fig.~\ref{OB} then $[{\cal O},Q\}= 0$ is equivalent
to the closed string cohomology condition.

To summarize, for bpz$( {\cal O}) ={\cal O}$ non-trivial elements in
the cyclic cohomology $HC^*$ can exist only if the closed string
cohomology is non-trivial. More precisely, if we consider $C_n\in
CC^n$ then the ghost number of $C_n$ is given by
\begin{equation}
|C_n|=\begin{cases}|\phi|-2-n\,,\qquad\hbox{disk has a closed string puncture}\cr
|{\cal{O}}|-n\,,\qquad \hbox{ disk has no puncture (Fig. \ref{3wd})}\end{cases}
\end{equation}
The first case corresponds to an insertion of an asymptotic on-shell
closed string sate, whereas the second corresponds to an insertion of
the form ${\cal{O}}|0\rangle_c$. In particular, $C_n \in HC^n$ with
ghost number $-n$ correspond either to a closed string asymptotic
state with ghost number $2$ or a closed string state of the form
${\cal{O}}|0\rangle_c$ with ghost number $0$ and vanishing
momentum. Since the semi-relative closed string cohomology at ghost
number zero and vanishing mass$^2$ contains only the vacuum (or
equivalently the identity operator ${\cal{O}}=I$) we then conclude
that all elements in the cyclic cohomology $HC^n$ at ghost number $-n$
must correspond to a physical closed string state This is the main
result of this paper.

For completeness we should mention that there are, in addition,
deformations of OSFT by open string operators with $C_3=0$ but
$C_2\neq 0$. However, the corresponding connected diagrams are
equivalent to those obtained by deformations of the $3$-vertex so that
we need not consider them separately.

%%%%%%%%%%%%%%%%%%%%%%%%%%%%%%%%%%%%%%%%%%%%%%%%%%%%%%%%%%%%%%%%%%%%%%
\subsection{Strips} 
%%%%%%%%%%%%%%%%%%%%%%%%%%%%%%%%%%%%%%%%%%%%%%%%%%%%%%%%%%%%%%%%%%%%%%

\begin{figure}
\begin{center}
\input{stripsC3.pstex_t}
\caption{Definition of $C^\eps_3$. The symbol on the left-hand
side represents $V^\eps_3$, the three-vertex with strips. The
first symbol on the right-hand side represents the three-vertex
without strips.}
\label{stripsC3}
\end{center}
\end{figure}
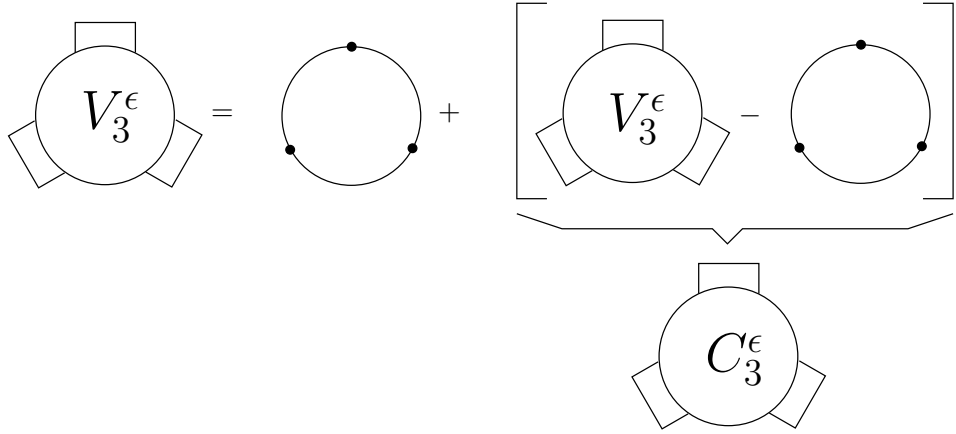

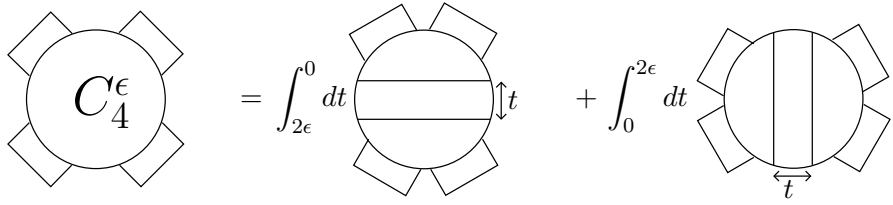
\begin{figure}
\begin{center}
\input{stripsC4.pstex_t}
\caption{Definition of $C^\eps_4$.}
\label{stripsC4}
\end{center}
\end{figure}

As a concrete example of the generic deformations discussed in the
last subsection we now describe a geometric deformation of the
3-vertex.  In Zwiebach's open-closed string field theory the vertices
in the action include strips of length $\pi$ for the external open
strings even in the absence of closed strings. This deformation should
not be physical since it is merely a reorganization of the moduli
space of the same theory. This has been shown in \cite{9301097} using
the BV formalism and also in \cite{Kajiura:2001ng} as a particular
case of $A_\infty$-quasi-isomorphisms. To see this in our context,
let us consider an infinitesimal version of Zwiebach's open-closed
theory in the absence of closed string insertions but with strips of
length $\eps>0$. We define $C^\eps_3$ pictorially in
Figure~\ref{stripsC3} as the deviation from Witten's vertex.
We can easily translate this picture to the algebraic expression
\begin{align}
C^\eps_3(\Psi_1, \Psi_2, \Psi_3) & =  \langle e^{-\eps L_0} \Psi_1, 
e^{-\eps L_0} \Psi_2, e^{-\eps L_0} \Psi_3 \rangle- 
\langle \Psi_1, \Psi_2, \Psi_3 \rangle \nn \\
&= -\eps \left( \langle L_0 \Psi_1, \Psi_2, \Psi_3 \rangle
+ \langle \Psi_1, L_0 \Psi_2, \Psi_3 \rangle + 
\langle \Psi_1, \Psi_2, L_0 \Psi_3 \rangle \right) + O(\eps^2).
\label{C3eps}
\end{align}
A four-vertex $C^\eps_4$ is now needed to produce the part of moduli
space of disks with four punctures on the boundary that is missed by
the Feynman diagrams constructed with $C^\eps_3$. The missing surfaces
are readily identified as the ones whose internal propagator has
length smaller than $2 \eps$. We can thus define $C^\eps_4$
pictorially as in Figure~\ref{stripsC4}. There are two ways to place
the propagator, hence the two contributions. Note that the integration
limits are reversed between the two contributions. This must be so in
order to parameterize smoothly the moduli space; we start with a
'vertical' propagator of length $2 \eps$, shrink it until is has
length zero, then we grow a 'horizontal' propagator until it has
length $2 \eps$.
We can again easily write the algebraic expressions of these diagrams
after we notice that the short propagator entering the definition of
$C^\eps_4$, has a very simple expression:
\begin{equation}
b_0 \, \int_0^{2 \eps}e^{-t L_0} dt = 2 \eps \, b_0 + O(\eps^2).
\end{equation}
We thus have
\begin{align}
C^\eps_4 &= 2 \epsilon \, (-1)^{\Psi_3+\Psi_4} \,
\langle \Psi_1 * \Psi_2, b_0 \, (\Psi_3 * \Psi_4) \rangle \nn \\
&\quad {} + 2 \epsilon \, (-1)^{\Psi_4+\Psi_1+\Psi_1 (\Psi_2+\Psi_3+\Psi_4)+1} \,
\langle \Psi_2 * \Psi_3, b_0 \, (\Psi_4 * \Psi_1) \rangle.
\label{C4eps}
\end{align} At $O(\eps)$ only $C^\eps_3$ and $C^\eps_4$ give
non-vanishing contribution since the volume of the moduli space of
$C^\eps_M$ is $O(\eps^{M-3})$. This can be seen by considering
tree-level diagrams involving $M\geq 3$ vertices. Let us now look at
the cohomology class of $(0,0,C^\eps_3, C^\eps_4, 0, \ldots)$. First,
it is straightforward to check that $C^\eps_3$ and $C^\eps_4$ are
cyclic. Next, $(0,0,C^\eps_3, C^\eps_4, 0, \ldots)$ should be closed;
let us check this explicitly.\footnote{It is not strictly necessary to
  check closedness since we will show later that this set of maps is
  exact, hence closed. Nevertheless, it is a good consistency check to
  go through this illustrative calculation.}  Since $Q$ commutes with
$L_0$, we find almost immediately that
\begin{equation}
0=(QC^\eps_3)(\Psi_1, \Psi_2, \Psi_3).
\label{closedLev3}
\end{equation}
The calculations of $(\delta C^\eps_3)$, $(Q C^\eps_4)$ and $(\delta
C^\eps_4)$ are not difficult, but it might be useful to recall one
more elementary identity that we need:
\begin{equation}
\langle {\cal O} A, B \rangle = (-1)^{{\cal O} A} \langle A, 
\text{bpz}({\cal O}) B \rangle.
\label{Identity4}
\end{equation}
It will be useful when the operator ${\cal O}$ is either $Q$, $L_0$ or
$b_0$, and we recall that $\text{bpz}(Q) = -Q$, $\text{bpz}(L_0) =
L_0$ and $\text{bpz}(b_0) = b_0$. It is then straightforward to show
that
\begin{equation}
(\delta C^\eps_3)(\Psi_1, \Psi_2,\Psi_3,\Psi_4) = 
2 \epsilon \, \left( \langle \Psi_1*\Psi_2, L_0(\Psi_3*\Psi_4) \rangle 
+ \langle \Psi_2*\Psi_3, L_0(\Psi_4*\Psi_1) \rangle \right).
\label{deltaC3eps}
\end{equation}
And the calculation of $QC^\eps_4$ is not much more difficult once
we realize that we must anticommute $Q$ past $b_0$ in order to cancel
out pairs of terms. The only remaining terms are then the ones
containing $\{Q, b_0\} = L_0$. Explicitly, we have
\begin{align}
(QC^\eps_4)(\Psi_1,\Psi_2,\Psi_3,\Psi_4) &= 2 \eps \,
\left((-1)^{\Psi_1+\Psi_2+\Psi_3+\Psi_4+1} \langle \Psi_1*\Psi_2,
  L_0 (\Psi_3*\Psi_4)
  \rangle \right. \nn \\
&\quad \left. + (-1)^{\Psi_1+\Psi_2+\Psi_3+\Psi_4 +
    \Psi_1(\Psi_2+\Psi_3+\Psi_4)} \langle \Psi_2*\Psi_3, L_0
  (\Psi_4*\Psi_1) \rangle \right).
\label{QC4eps}
\end{align}
Now, in order to compare (\ref{deltaC3eps}) and (\ref{QC4eps}), we
must have recourse to the fact that, by ghost-number counting, they
are nonzero only when $|\Psi_1| + |\Psi_2| + |\Psi_3| + |\Psi_4| =
3$. We can then conclude that 
\begin{equation}
\delta C^\eps_3 - QC^\eps_4 = 0.
\label{closedLev4}
\end{equation}
At last, it is also straightforward to check that 
\begin{equation}
\delta C^\eps_4 = 0.
\label{closedLev5}
\end{equation}
And we can combine (\ref{closedLev3}), (\ref{closedLev4}) and
(\ref{closedLev5}) into the closedness relation
\begin{equation}
(\delta-(-1)^N Q) \, (0, 0, C^\eps_3, C^\eps_4, 0, \ldots) = 0.
\end{equation}

Let us now show that $(0, 0, C^\eps_3, C^\eps_4, 0, \ldots)$ is an
exact element of the cyclic cohomology. To this end we will try to
find a $D^\eps_3$ such that $Q D^\eps_3 = C^\eps_3$ and $\delta
D^\eps_3 = C^\eps_4$. It turns out that we will be lucky, and we will
need neither a $D^\eps_1$ nor a $D^\eps_2$. It remains to write down
an expression for $D^\eps_3$; but if we focus on the requirement that
$Q D^\eps_3 = C^\eps_3$ and on the expression (\ref{C3eps}) for
$C^\eps_3$, we see that a natural candidate is simply the expression
of $C^\eps_3$ with all $L_0$'s replaced by $b_0$'s. This is almost
right, we just need to tweak some signs. We claim that the expression
for $D^\eps_3$ is
\begin{equation}
D^\eps_3(\Psi_1, \Psi_2, \Psi_3) \equiv -\eps \, \left( 
\langle b_0 \Psi_1, \Psi_2, \Psi_3 \rangle + 
(-1)^{\Psi_1} \langle \Psi_1, b_0 \Psi_2, \Psi_3 \rangle + 
(-1)^{\Psi_1+\Psi_2} \langle \Psi_1, \Psi_2, b_0 \Psi_3 \rangle \right).
\label{D3}
\end{equation}
We can show that 
\begin{equation}
Q D^\eps_3 = C^\eps_3.
\label{QD3}
\end{equation}
We sketch the straightforward proof by acting with $Q$ on the first
term in the parentheses.  We have
\begin{align*}
& \langle b_0 Q \Psi_1, \Psi_2, \Psi_3 \rangle + (-1)^{\Psi_1} \langle 
b_0 \Psi_1, Q \Psi_2, \Psi_3 \rangle + (-1)^{\Psi_1 + \Psi_2} \langle
b_0 \Psi_1, \Psi_2, Q \Psi_3 \rangle = \\
& = \langle L_0 \Psi_1, \Psi_2, \Psi_3 \rangle - (-1)^{\Psi_1} \langle 
b_0 \Psi_1, Q(\Psi_2*\Psi_3) \rangle \\
& \hspace{3.0ex} + (-1)^{\Psi_1} \langle b_0 \Psi_1, (Q \Psi_2)*\Psi_3 \rangle
+ (-1)^{\Psi_1+\Psi_2} \langle b_0 \Psi_1, \Psi_2 *(Q \Psi_3) \rangle \\
& = \langle L_0 \Psi_1, \Psi_2, \Psi_3 \rangle,
\end{align*}
and the cancellation of terms containing $Q$ works in the same way for
the other two terms in Eq.~(\ref{D3}). This establishes (\ref{QD3}).
Furthermore, we claim that 
\begin{equation}
(\delta D^\eps_3)(\Psi_1, \Psi_2, \Psi_3, \Psi_4) = 
C^\eps_4(\Psi_1, \Psi_2, \Psi_3, \Psi_4).
\label{deltaQ3}
\end{equation}
The proof is mechanical. In short, there are two kinds of terms in
$\delta D^\eps_3$: Terms containing $b_0 \Psi_i$ and terms containing
$b_0(\Psi_i * \Psi_j)$. All the terms of the first kind cancel out by
pairs. And the four terms of the second kind can be grouped into two
terms (with a factor 2 in front) by virtue of Eq.~(\ref{Identity4}),
for instance
$$
\langle b_0(\Psi_1*\Psi_2), \Psi_3, \Psi_4 \rangle = 
\langle b_0(\Psi_1*\Psi_2), \Psi_3*\Psi_4 \rangle = 
(-1)^{\Psi_1+\Psi_2} \langle \Psi_1*\Psi_2, b_0(\Psi_3*\Psi_4) \rangle.
$$
At last we use the fact that everything trivially vanishes unless
$|\Psi_1|+|\Psi_2|+|\Psi_3|+|\Psi_4| = 4$, and we get precisely the
same expression as in Eq.~(\ref{C4eps}). Therefore Eq.~(\ref{deltaQ3})
is verified. Grouping (\ref{QD3}) and (\ref{deltaQ3}), we can write
\begin{equation}
(0,0,C^\eps_3, C^\eps_4, 0, \ldots) = (\delta - (-1)^N Q) \, (0,0,D^\eps_3, 0, 0, \ldots),
\end{equation}
which means that the geometric deformations do not contribute to the
cyclic cohomology.

%%%%%%%%%%%%%%%%%%%%%%%%%%%%%%%%%%%%%%%%%%%%%%%%%%%%%%%%%%%%%%%%%%%%%%
\section{Background Independence}\label{bi}
%%%%%%%%%%%%%%%%%%%%%%%%%%%%%%%%%%%%%%%%%%%%%%%%%%%%%%%%%%%%%%%%%%%%%%

Let us expand the theory around a classical solution of the equations
of motion, $\Psi \rightarrow \Psi_0+\Psi$ with $Q\Psi_0 + \Psi_0 *
\Psi_0 = 0$. The Witten action is then the same except for the
appearance of a cosmological constant (irrelevant for us) and for the
fact that $Q$ is replaced by a new BRST operator $Q'$ given by
\begin{equation}
Q'\Psi = Q\Psi + \Psi_0*\Psi + (-1)^{1+\Psi} \Psi*\Psi_0.
\end{equation}
We claim that the cyclic cohomology of $(\delta -(-1)^N Q')$ is
isomorphic to the cyclic cohomology of $(\delta -(-1)^N Q)$.  Our
strategy will be to find a linear bijection $h$
\begin{equation}
h : \prod_{n=1}^\infty \text{Hom}(A^n, \mathbb{R}) \ \longrightarrow 
\ \prod_{n=1}^\infty \text{Hom}(A^n, \mathbb{R})
\end{equation}
such that 
\begin{equation}
(\delta -(-1)^N Q') \, h(C_1, C_2, \ldots) = 0 \quad \text{iff} \quad
(\delta -(-1)^N Q)  \, (C_1, C_2, \ldots) = 0,
\end{equation} 
and such that $h(C_1, C_2, \ldots)$ is $(\delta -(-1)^N Q')$-exact if
and only if $(C_1, C_2, \ldots)$ is $(\delta -(-1)^N Q)$-exact, where
$C_n \in \text{Hom}(A^n, \mathbb{C})$.  If we denote $h(C_1, C_2,
\ldots)$ by $(D_1, D_2, \ldots)$, our candidate for $h$ is
\begin{equation}
D_n(\Psi_1, \ldots, \Psi_n) = 
\sum_{k=0}^\infty (-1)^{(n+1) k + \frac{1}{2} k (k+1)}
\sum_{f \in {\cal F}^n_k} (-1)^{\sum_{i=1}^n (\Psi_i-1) (k+f(i)-i)} 
C_{n+k}^f(\Psi_1, \ldots, \Psi_n),
\label{h}
\end{equation}
where ${\cal F}^n_k$ is the set of strictly increasing functions from
$\{1, \ldots, n\}$ into $\{1, \ldots, n+k\}$ such that $f(n) = n+k$;
and $C_{n+k}^f(\Psi_1, \ldots, \Psi_n)$ is an element of
$\text{Hom}(A^n, \mathbb{C})$ defined by
\begin{equation}
C_{n+k}^f(\Psi_1, \ldots, \Psi_n) = 
C_{n+k}(\Psi_0,\ldots,\Psi_0,\Psi_1, \Psi_0,\ldots,\Psi_0,\Psi_2,\ldots 
\ldots,\Psi_n),
\end{equation}
where the $\Psi_i$'s are inserted at the positions $f(i)$, and the
remaining $k$ slots are filled with $\Psi_0$'s. Note that we may have
one or more (or none) $\Psi_0$ on the left of $\Psi_1$, but there
cannot be any on the right of $\Psi_n$. In this way we avoid
over-counting since a $\Psi_0$ on the right of $\Psi_n$ can be moved
to the left of $\Psi_1$ by cyclicity.

Using the fact that $D_n(\Psi_1, \ldots, \Psi_n)$ vanishes unless 
$(-1)^{\Psi_1 + \ldots +\Psi_n} = (-1)^n$ we readily check that $D_n$ is cyclic, namely
\begin{equation}
D_n(\Psi_2, \Psi_3, \ldots, \Psi_n, \Psi_1) = 
(-1)^{\Psi_1 (\Psi_2 + \ldots + \Psi_n) + n+1} D_n(\Psi_1, \Psi_2, \ldots, \Psi_n).
\end{equation}
It is also straightforward (although a bit lengthy) to check that
\begin{align}
& \delta D_{n-1} - (-1)^n Q' D_n = \nonumber \\
& \sum_{k=0}^\infty (-1)^{(n+1) k + \frac{1}{2} k (k+1)}
\sum_{f \in {\cal F}^n_k} (-1)^{\sum_{i=1}^n (\Psi_i-1) (k+f(i)-i)} 
\left(\delta C_{n+k-1} + (-1)^{n+k+1} Q C_{n+k} \right)^f.
\end{align}
But we recognize that the last line is precisely our function $h$
defined in Eq.~(\ref{h}), applied on $(\delta-(-1)^NQ)(C_1, C_2,
\ldots)$. We have thus found that
\begin{equation}
(\delta - (-1)^N Q') \, h = h \, (\delta - (-1)^N Q).
\label{QQprime}
\end{equation}
This equation tells us immediately that $h$ maps $(\delta - (-1)^N
Q)$-closed elements to $(\delta - (-1)^N Q')$-closed elements. It also
tells us that $h$ maps $(\delta - (-1)^N Q)$-exact elements to
$(\delta - (-1)^N Q')$-exact elements. In order to conclude that the
two cyclic cohomologies are isomorphic, we now just need to show that
$h$ is invertible, so that we could write
\begin{equation}
(\delta - (-1)^N Q) \, h^{-1} = h^{-1} \, (\delta - (-1)^N Q').
\label{QQprimeinverse}
\end{equation}
Indeed, (\ref{QQprime}) and (\ref{QQprimeinverse}) show that $h$ is a
linear bijection from $\text{Ker}(\delta - (-1)^N Q)$ onto
$\text{Ker}(\delta - (-1)^N Q')$ and is also a linear bijection from
$\text{Im}(\delta - (-1)^N Q)$ onto $\text{Im}(\delta - (-1)^N Q')$.

Let us now show that $h$ is invertible. First, we observe that 
\begin{equation}
Q'\Psi_0 = Q \Psi_0 + \Psi_0 * \Psi_0 + \Psi_0 * \Psi_0 = \Psi_0 * \Psi_0,
\end{equation}
in other words $(-\Psi_0)$ is a solution of the equations of motion
with respect to $Q'$. We note also that 
\begin{equation}
Q \Psi = Q' \Psi + (-\Psi_0)*\Psi + (-1)^{1+\Psi} \Psi* (-\Psi_0).
\end{equation}
Since the only facts we used in order to construct $h$ were that
$\Psi_0$ obeys the equations of motion with respect to $Q$ and that
$Q' \Psi = Q \Psi + \Psi_0*\Psi + (-1)^{1+\Psi} \Psi*\Psi_0$, we see
that the inverse map $h^{-1}$ must be given by Eq.~(\ref{h}) with
$\Psi_0$ replaced with $-\Psi_0$.  Since $\Psi_0$ appears $k$ times in
the sum on the right-hand side of (\ref{h}), this amounts to simply
introducing a factor of $(-1)^k$. Explicitly, we have therefore
\begin{equation}
C_n(\Psi_1, \ldots, \Psi_n) = 
\sum_{k=0}^\infty (-1)^{nk + \frac{1}{2} k (k+1)}
\sum_{f \in {\cal F}^n_k} (-1)^{\sum_{i=1}^n (\Psi_i-1) (k+f(i)-i)} 
D_{n+k}^f(\Psi_1, \ldots, \Psi_n).
\end{equation}
This concludes our proof that the cyclic cohomology is background
independent.

%%%%%%%%%%%%%%%%%%%%%%%%%%%%%%%%%%%%%%%%%%%%%%%%%%%%%%%%%%%%%%%%%%%%%%
\section{Conclusions}\label{conc}
%%%%%%%%%%%%%%%%%%%%%%%%%%%%%%%%%%%%%%%%%%%%%%%%%%%%%%%%%%%%%%%%%%%%%%

In this paper we showed that the closed string cohomology is
isomorphic to the cyclic cohomology of cubic open string field
theory. The latter is defined solely in terms of the structure of open
string theory, i.e. the open string BRST operator and the product on
the algebra of open string fields. Part of this result could have been
anticipated since a generic field theory in flat space, formulated in
a covariant form admits an essentially unique linearized coupling to
gravity in terms of the Noether procedure. What is special about
string theory, however, is that on top of that, open string theory
knows about the on-shell condition of the closed string states in
terms of cyclic cohomology.

The natural question that arises then is whether there is more
information to be uncovered by considering second order deformations
in the closed string deformation. We think the answer is positive
since a general property of the cohomology ring $HC(A)$ is that it
possesses a natural Lie super-algebra structure \cite{Penkava:1994mu},
more precisely an $L_\infty$-structure just like closed string field
theory. So one can hope to learn more on closed strings by considering
consistent deformations of OSFT. On the other hand there are certain
obstructions in extending infinitesimal deformations to second
order. This is, however, expected since not every marginal closed
string deformation is exactly marginal.

Are there any applications of our results to open string field theory?
For one thing the reformulation of closed string cohomology and in
particular the open closed vertices in terms of cyclic cohomology of
OSFT together with the background independence of the latter provides
a formal definition of linearized open-closed string theory in any
open string background. For instance, in Schnabl's vacuum solution
where the open string cohomology is empty but the closed string
cohomology is not we conclude that the closed string spectrum is still
encoded in OSFT.

Finally we would like to mention another possible avenue for further
investigation. Within boundary string field theory it has been argued
\cite{Baumgartl:2004iy} that certain deformations of the closed string
background are equivalent to "collective excitations" of the open
string (i.e. insertions of non-local boundary interactions on the
boundary of the world sheet). At first sight this seems to be in
contradiction with our present result that there is an isomorphism
between non-trivial deformations of OSFT and closed string insertions
in the bulk of the disk. However, we should note that the argument
given in section \ref{un} is based on insertion of generic local
operators on the boundary. It would be interesting to relax the latter
condition.

\section*{Acknowledgments}
We would like to thank M.~Kiermaier for collaboration at initial
stages of this work, and M.~Schnabl, A.~Sen and B.~Zwiebach for
helpful discussions, as well as K.~Cieliebak for explaining to us the
relation to algebraic topology in appendix \ref{HH}. This work was
supported in part by the Transregional Collaborative Research Centre
TRR 33, the DFG cluster of excellence ``Origin and Structure of the
Universe'' as well as the DFG project Ma 2322/3-1.

%%%%%%%%%%%%%%%%%%%%%%%%%%%%%%%%%%%%%%%%%%%%%%%%%%%%%%%%%%%%%%%%%%%%%%
\appendix
\section{Geometric Interpretation}\label{HH}
%%%%%%%%%%%%%%%%%%%%%%%%%%%%%%%%%%%%%%%%%%%%%%%%%%%%%%%%%%%%%%%%%%%%%%

In order to gain a geometric intuition of the $A_\infty$-algebra
arising in our context and Hochschild cohomology in particular it may
be helpful to draw an analogy with algebraic topology. The relation of
the closed string cohomology to the de Rham cohomology in the loop
space will be discussed in subsection \ref{closedLoops}.

%%%%%%%%%%%%%%%%%%%%%%%%%%%%%%%%%%%%%%%%%%%%%%%%%%%%%%%%%%%%%%%%%%%%%%
\subsection{$A_\infty$-algebras}
%%%%%%%%%%%%%%%%%%%%%%%%%%%%%%%%%%%%%%%%%%%%%%%%%%%%%%%%%%%%%%%%%%%%%%

Let $M$ be a smooth manifold and $(A,\de)=(\Omega^*(M),\de)$ its de
Rham complex. Then $(A,\de,\wedge)$ defines a differential graded
algebra with the operations
\begin{eqnarray}
\de&:& A\to A\nonumber\\
&&x\mapsto \de x\nonumber\\
\wedge&:&A\otimes A\to A\nonumber\\
&&(x,y)\mapsto x\wedge y
\end{eqnarray}
The exterior derivative $\de$ with $\de^2=0$ has degree $|\de|=1$
while $\wedge$ has degree $|\wedge|=0$. An $A_\infty$-algebra is
obtained as a certain deformation of $(A,\de,\wedge)$ including higher
maps $ A^{\otimes n}\to A$. In order to do this it is convenient to
have a uniform grading for these maps. This can be achieved as
follows. We say the $x\in \Omega^{p}$ has degree $p$. Then $\de x\in
\Omega^{p+1}$ has degree $p+1$ and $x\wedge y\in\Omega^{p+q}$ has
degree $p+q$. We then define the grading of $x$ as
$\mathrm{grad}(x)=|x|-1$, so that $\mathrm{grad}
(\de)=\mathrm{grad}(\wedge)=1$.  This is called a shift (or a
suspension) and we denote the shifted vector space by $A[1]$. In
particular,
\begin{eqnarray}
b_1(x)= \de x&:& A[1]\to A[1]\nonumber\\
b_2(x, y)=(-1)^{|x|+1} x\wedge y&:&A[1]\otimes A[1]\to A[1]\nonumber\\
\end{eqnarray}
have both grading 1. $A_\infty$-deformations of $(A[1],b_1, b_2)$ are
obtained as follows (see e.g. \cite{GJ}): We define the Bar complex
\begin{equation}
BA[1]=\oplus_{k\geq 1}A[1]^{\otimes k}=\oplus_k B_kA[1]
\end{equation}
and continuations $\hat{b}_i:BA[1]\to BA[1]$, $i=1,2$ as coderivatives
with components
\begin{equation}
(\hat b_i)_{k,k-i}=\sum\limits_{t = 0}^{k-i} {\bf 1}^{\otimes k-i-t }\otimes 
b_i\otimes {\bf 1}^{\otimes t}
\end{equation}
It is not hard to see that $\hat{b}=\hat{b}_1+\hat{b}_2$ squares to
zero and hence $(BA[1],\hat{b})$ is a cochain complex. The
$A_\infty$-generalization of $(BA[1],\hat{b})$ is obtained by
including higher maps $ b_s :A[1]^{\otimes s}\to A[1], s\geq 1$ so
that the coderivative with components
\begin{equation}
  (\hat {b})_{n,u}=\sum\limits_{\substack{r+s+t = n\\r+1+t = u}} 
  {\bf 1}^{\otimes r}\otimes b_s\otimes  {\bf 1}^{\otimes t}
\end{equation}
squares to zero. Now, if $\hat b$ is a coderivative of grade $1$ on
$BA[1]$ with components $b_n : A[1]^{\otimes n} \to A$ , then its
square is a coderivative of grade $2$ with components
\begin{equation}
(\hat b^2)_n=\sum\limits_{i+j = n+1}\sum\limits_{k=0}^{n-j}b_i\cdot( 
{\bf 1}^{\otimes k}\otimes b_j\otimes  {\bf 1}^{\otimes n-k-j})
\end{equation}
Imposing that $\hat b^2=0$, we obtain a characterization of all
differentials compatible with the $A_\infty$-structure on $BA[1]$. For
example,
\begin{align}
b_1\cdot b_2&+b_2\cdot(b_1\otimes {\bf 1}+ {\bf 1}\otimes b_1)=0\,,\qquad & n=2 \nn \\
b_1 \cdot b_3 + 
b_3\cdot (b_1\otimes {\bf 1}\otimes {\bf 1})+b_3\cdot( {\bf 1}\otimes b_1\otimes {\bf 1}) &+ 
b_3\cdot( {\bf 1}\otimes {\bf 1}\otimes b_1) \nn \\
\hspace{7cm} &+b_2\cdot(b_2\otimes {\bf 1}+ {\bf 1}\otimes b_2)=0\,,\qquad & n=3
\end{align}
We can translate these relations to the maps $m_k :A^{\otimes k}\to A$
without the shift. For this we recall that if $A$ is a graded vector
space over ${\mathbb C}$ then its suspension, $A[1]$ is the graded
${\mathbb C}$-vector space with a grading shifted by $1$. We define a
shift operator $s$ as $A[1]=sA$, or
\begin{equation}
(sA)_i=A_{i-1}.
\end{equation}
Now, if $b_k : (sA)^{\otimes k} \to sA$ is a multilinear map of
grading $1$ one defines $m_k : A^{\otimes k} \to A$ of degree
$|m_k|=2-k$ by\footnote{Since the signs depend only on the grading 
modulo ${\mathbb Z}_2$, one could simplify the notation by writing $s^{-1}=s$.}
\begin{equation}
b_k=s \cdot m_k\cdot {\left(s^{-1}\right)}^{\otimes k}
\end{equation}
or equivalently
\begin{equation}\label{bm}
b_k(sa_1, \cdots sa_k)=(-1)^{(k-1)|a_1|+(k-2)|a_2|+\cdots+2|a_{k-2}|+|a_{k-1}|
+k(k-1)/2} s m_k(a_1, \cdots a_k)
\end{equation}
We display the first few of the $m_k$'s explicitly 
\begin{align}
b_1(sa)&=s m_1(a)\\
b_2(sa_1,sa_2)&=(-1)^{|a_1|+1} s m_2(a_1,a_2)\nn\\
b_3(sa_1,sa_2,sa_3)&=(-1)^{2|a_1|+|a_2|+3} s m_3(a_1,a_2,a_3)\nn
\end{align}
which in turns leads to
\begin{equation}
m_1(m_2(a_1,a_2))-m_2(m_1(a_1),a_2)-(-1)^{|a_1|} m_2(a_1,m_1(a_2)) = 0,
\end{equation}
and
\begin{align}\label{ai}
m_1(m_3(a_1,a_2,a_3))+m_3(m_1(a_1),a_2,a_3)+(-1)^{|a_1|} m_3(a_1,m_1(a_2),a_3)&\nn\\
+(-1)^{|a_1|+|a_2|} m_3(a_1,a_2,m_1(a_3))-m_2(m_2(a_1,a_2),a_3)+ m_2(a_1,m_2(a_2,a_3))&=0,
\end{align}
and so on. In particular, $m_2$ is associative only up to the higher
homotopy map $m_3$.

The infinitesimal $A_\infty$-deformations of the DGA of OSFT relevant
to our work are described in the next subsection.

%%%%%%%%%%%%%%%%%%%%%%%%%%%%%%%%%%%%%%%%%%%%%%%%%%%%%%%%%%%%%%%%%%%%%%
\subsection{Hochschild Cohomology}
%%%%%%%%%%%%%%%%%%%%%%%%%%%%%%%%%%%%%%%%%%%%%%%%%%%%%%%%%%%%%%%%%%%%%%

Let us now consider an infinitesimal $A_\infty$-deformation of the
differential graded algebra $(BA[1],\hat{\bar b})$. For that we write
\begin{eqnarray}
b_1&=&\bar b_1+\epsilon {\cal{F}}_1\nn\\
b_2&=&\bar b_2+\epsilon {\cal{F}}_2\\
b_k&=&\epsilon {\cal{F}}_k\;\qquad k>2\nonumber 
\end{eqnarray}
The $A_\infty$-condition $\hat b^2=0$ then implies that
\begin{equation}
\de_H(\hat {\cal{F}})=\{\hat{\bar b},\hat {\cal{F}}\}=0
\end{equation}
where $\de_H(\hat {\cal{F}})$ is the Hochschild co-boundary operator
and $\de_H^2(\hat {\cal{F}})\equiv 0$ follows from the Jacobi identity
for $\{,\}$. We denote the corresponding complex by
$CH^*(A[1],\de_H)$.

We can again translate these formulas to deformations of $m_k
:A^{\otimes k}\to A$ without the shift, that is
\begin{align}
m_1&=\de+\eps f_1\nonumber\\
m_2&= \wedge-\eps f_2\nonumber\\
m_k&= (-1)^{k+1}\eps f_k\,,\qquad k>2
\end{align}
we find that at order $\eps$
\begin{equation}\label{h1}
((-1)^{|f|}\de-\delta)\hat f=0
\end{equation}
where 
\begin{eqnarray}
\delta f_k (x_1,\cdots.x_{k+1})&=&(-1)^{|x_1||f_k|} x_1\wedge f_k(x_2,\cdots.x_{k+1})\\
&&+\sum\limits_{i=0}^k(-1)^i f_k(x_1,\cdots,x_i\wedge x_{i+1}\cdots,x_{k+1})\nn\\
&&+(-1)^{n+1}f_k(x_1,\cdots.x_{k})\wedge x_{k+1}\,.\nn
\end{eqnarray}

%%%%%%%%%%%%%%%%%%%%%%%%%%%%%%%%%%%%%%%%%%%%%%%%%%%%%%%%%%%%%%%%%%%%%%
\subsection{Closed Loops} \label{closedLoops}
%%%%%%%%%%%%%%%%%%%%%%%%%%%%%%%%%%%%%%%%%%%%%%%%%%%%%%%%%%%%%%%%%%%%%%

So far we have discussed the $A_\infty$-deformations of the cochain
complex $(BA[1],\hat{b})$ which is the geometric analog of the
differential graded algebra of open string field theory. Next we would
like to draw an analogy with closed string insertions. To illustrate
this relations we will consider the chain complex $(C_*(M),\partial)$
together with the cap product $\cap$, although this does not strictly
speaking provide an algebra. Ignoring this fact for the time being we
then consider the product
\begin{equation}
C_*(\Lambda_0M)\times C_*(M)^n
\end{equation}
where $\Lambda_0M$ is the space of loops in $M$ with fixed base
point. We can then define a map
\begin{equation}
\varphi: C_*(\Lambda_0M)\times C_*(M)^n\to C_*(M)
\end{equation}
as follows. Let $g_i$ and $h$ be homeomorphisms from a standard
simplex $P$ to $M$ and $\Lambda_0M$ respectively,
\begin{eqnarray}
g_i:( \sigma\in P)&&\mapsto g_i(\sigma)\in M\\
h:(\tau\in P,t\in[0,1])&&\mapsto h(\tau,t)\in M
\end{eqnarray}
For fixed $t_1<t_2<\cdots<t_n$ the intersection
\begin{equation}
\begin{cases}g_1(\sigma_1)&=h(\tau,t_1)\cr
g_2(\sigma_2)&=h(\tau,t_2)\cr
\cdots\cr
g_n(\sigma_n)&=h(\tau,t_n)
\end{cases}
\end{equation}
defines an element in $ C_*(M)$ 
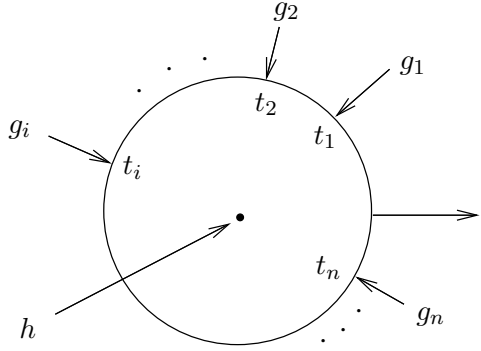
\begin{figure}
\begin{center}
\input{loops.pstex_t}
\caption{Illustration of the map $\varphi$.}
\label{loops}
\end{center}
\end{figure}
that can be represented in terms of a disc diagram as in
Fig.~\ref{loops} and which generalizes the (transverse) intersection
in $ C_*(M)$. To complete this construction we should integrate over
$0\leq t_1<t_2<\cdots<t_n<1$. Ignoring the issues about transversality
the corresponding map then induces a chain map
\begin{equation}
\varphi : C_*(\Lambda_0M)\to CH_*(C_*(M,\partial,\cap), \de_H)\;;\qquad 
\de_H\varphi(c)=\varphi(\partial c)\;,\quad c\in   C_*(\Lambda_0M)
\end{equation}
where $\de_H$ is the coboundary operator on the space
$CH_*(C_*(M),\partial,\cap)$ of deformations of
$(C_*(M),\partial,\cap)$.

Returning again to cochains on $M$ one can make a precise statement:
The method of Chen's iterated integrals over $0\leq
t_1<t_2<\cdots<t_n<1$ (see e.g. \cite{Fukaya}) provides an isomorphism
between the homology group of $\Lambda_0 M$ and the Hochschild
cohomology of $(\Omega^*(M),\de,\wedge)$
\begin{equation}
I_*:H_*(\Lambda_0M)\stackrel{\cong}{\to}HH^*(\Omega^*(M),\de,\wedge)
\end{equation}
This is the analog, in algebraic topology, of the isomorphism derived
in section \ref{GD}.

%%%%%%%%%%%%%%%%%%%%%%%%%%%%%%%%%%%%%%%%%%%%%%%%%%%%%%%%%%%%%%%%%%%%%%
\subsection{Cyclic Cohomology}\label{hccw}
%%%%%%%%%%%%%%%%%%%%%%%%%%%%%%%%%%%%%%%%%%%%%%%%%%%%%%%%%%%%%%%%%%%%%%

In addition to specifying the data $(A,Q,*)$, the construction of
string field theory requires an invariant inner product
$\langle.\,,.\rangle$ on $A$, i.e. $\langle a*b, c\rangle = \langle a,
b*c \rangle$ and $\langle a, b\rangle = (-1)^{|a||b|} \langle b, a
\rangle$. Note also that $\langle a, b \rangle$ is non-vanishing only
if $|a|+|b|=3$. Deformations of $A$ preserving this inner product are
governed by cyclic cohomology. To see this connection we note that in
the presence of an invariant inner product, there is a natural
isomorphism $\mathrm{Hom}(A^M,A)\to \mathrm{Hom}(A^{M+1},{\mathbb
  C})$. The image of $f_{M}\in \mathrm{Hom}(A^{M},A)$ in
$\mathrm{Hom}(A^{M+1},{\mathbb C})$ is then
\begin{equation}
C_{M+1}(f)(\Psi_1,\cdots,\Psi_{M+1})= \langle f_{M}(\Psi_1,\cdots,\Psi_{M}),\Psi_{M+1}\rangle
\end{equation}
The cyclic symmetry (\ref{cycl}) implies, in particular, that
$f_2(a,b)$ preserves the inner product since
\begin{equation}
\langle f_2(a,b),c\rangle= C_3(a,b,c)= (-1)^{|a|(|b|+|c|)}C_3(b,c,a)= 
(-1)^{|a|(|b|+|c|)} \langle f_2(b,c),a\rangle= \langle a,f_2(b,c) \rangle
\end{equation}
The Hochschild co-boundary operator $\delta$ on $\hbox{Hom}(A^M,A)$
induces a co-boundary operator $\delta$ on $\hbox{Hom}(A^M,{\mathbb
 C})$ through $\delta C_M(f)= C_M(\delta f)$. This gives
\begin{eqnarray}
(\delta C_M)(\Psi_1,\cdots,\Psi_{M+1})&=&\sum\limits_{i=1}^{M}
(-1)^iC_M(\Psi_1,\cdots,\Psi_i\wedge\Psi_{i+1},\cdots,\Psi_{M+1})\\
&&+(-1)^{\Psi_1(\Psi_2+\cdots+\Psi_{M+1})}C_M(\Psi_2,\cdots,\Psi_M,\Psi_{M+1}\wedge\Psi_1)\nn
\end{eqnarray}
This co-boundary operator takes cyclic elements to cyclic elements. The action of the differential $Q$ on $CC^*(A)$ is defined by
\begin{equation}
(Q C_M)(\Psi_1, \ldots, \Psi_M) = \sum_{i=1}^M (-1)^{\Psi_1 + \ldots + \Psi_{i-1}}
C_M(\Psi_1, \ldots, Q \Psi_i, \ldots, \Psi_M).
\end{equation}
It also takes cyclic elements to cyclic elements. Since $Q^2 = \delta^2 = [Q, \delta] = 0$, the operator $(\delta - (-1)^M Q)$ squares to zero. 
Now, if we
denote the submodule of $\hbox{Hom}(A^{M},{\mathbb C})$ consisting of
cyclic elements by $CC^M(A)$, then the cyclic cohomology of $A$ is
defined by
\begin{equation}
HC^*(A) = \mathrm{ker}[(\delta - (-1)^M Q): CC^*(A)\to CC^*(A)]/
 \mathrm{Im}[(\delta - (-1)^M Q): CC^*(A)\to CC^*(A)].
\end{equation}
This is the cyclic cohomology that we argue to be isomorphic to the 
cohomology of closed strings.

%%%%%%%%%%%%%%%%%%%%%%%%%%%%%%%%%%%%%%%%%%%%%%%%%%%%%%%%%%%%%%%%%%%%%%%%%%%%%%

\end{document}

%% file: ocdisk.pstex_t
\begin{picture}(0,0)%
\includegraphics{ocdisk.pstex}%
\end{picture}%
\setlength{\unitlength}{4144sp}%
\begingroup\makeatletter\ifx\SetFigFontNFSS\undefined%
\gdef\SetFigFontNFSS#1#2#3#4#5{%
  \reset@font\fontsize{#1}{#2pt}%
  \fontfamily{#3}\fontseries{#4}\fontshape{#5}%
  \selectfont}%
\fi\endgroup%
\begin{picture}(2963,2674)(1651,-6524)
\put(2295,-5988){\makebox(0,0)[lb]{\smash{{\SetFigFontNFSS{11}{13.2}{\familydefault}{\mddefault}{\updefault}$.$}}}}
\put(2211,-5864){\makebox(0,0)[lb]{\smash{{\SetFigFontNFSS{11}{13.2}{\familydefault}{\mddefault}{\updefault}$.$}}}}
\put(2140,-5708){\makebox(0,0)[lb]{\smash{{\SetFigFontNFSS{11}{13.2}{\familydefault}{\mddefault}{\updefault}$.$}}}}
\put(2113,-5539){\makebox(0,0)[lb]{\smash{{\SetFigFontNFSS{11}{13.2}{\familydefault}{\mddefault}{\updefault}$.$}}}}
\put(2094,-5358){\makebox(0,0)[lb]{\smash{{\SetFigFontNFSS{11}{13.2}{\familydefault}{\mddefault}{\updefault}$.$}}}}
\put(4045,-6272){\makebox(0,0)[lb]{\smash{{\SetFigFontNFSS{11}{13.2}{\familydefault}{\mddefault}{\updefault}$.$}}}}
\put(4200,-6161){\makebox(0,0)[lb]{\smash{{\SetFigFontNFSS{11}{13.2}{\familydefault}{\mddefault}{\updefault}$.$}}}}
\put(4409,-5881){\makebox(0,0)[lb]{\smash{{\SetFigFontNFSS{11}{13.2}{\familydefault}{\mddefault}{\updefault}$.$}}}}
\put(4307,-6027){\makebox(0,0)[lb]{\smash{{\SetFigFontNFSS{11}{13.2}{\familydefault}{\mddefault}{\updefault}$.$}}}}
\put(3868,-6388){\makebox(0,0)[lb]{\smash{{\SetFigFontNFSS{11}{13.2}{\familydefault}{\mddefault}{\updefault}$.$}}}}
\put(2891,-4612){\makebox(0,0)[lb]{\smash{{\SetFigFontNFSS{11}{13.2}{\familydefault}{\mddefault}{\updefault}$b$}}}}
\put(2937,-6038){\makebox(0,0)[lb]{\smash{{\SetFigFontNFSS{11}{13.2}{\familydefault}{\mddefault}{\updefault}$b$}}}}
\put(2558,-5119){\makebox(0,0)[lb]{\smash{{\SetFigFontNFSS{11}{13.2}{\familydefault}{\mddefault}{\updefault}$b$}}}}
\put(4125,-5258){\makebox(0,0)[lb]{\smash{{\SetFigFontNFSS{11}{13.2}{\familydefault}{\mddefault}{\updefault}$b$}}}}
\put(3326,-5128){\makebox(0,0)[lb]{\smash{{\SetFigFontNFSS{11}{13.2}{\familydefault}{\mddefault}{\updefault}$\phi(0)$}}}}
\put(4599,-5118){\makebox(0,0)[lb]{\smash{{\SetFigFontNFSS{11}{13.2}{\familydefault}{\mddefault}{\updefault}$\Psi^{h_{M-1}}_{M-1}(w_{M-1})$}}}}
\put(4240,-4413){\makebox(0,0)[lb]{\smash{{\SetFigFontNFSS{11}{13.2}{\familydefault}{\mddefault}{\updefault}$\Psi^{h_M}_M(w_M)$}}}}
\put(1666,-4831){\makebox(0,0)[lb]{\smash{{\SetFigFontNFSS{11}{13.2}{\familydefault}{\mddefault}{\updefault}$\Psi^{h_2}_2(w_2)$}}}}
\put(2341,-6451){\makebox(0,0)[lb]{\smash{{\SetFigFontNFSS{11}{13.2}{\familydefault}{\mddefault}{\updefault}$\Psi^{h_i}_i(w_i)$}}}}
\put(2791,-4021){\makebox(0,0)[lb]{\smash{{\SetFigFontNFSS{11}{13.2}{\familydefault}{\mddefault}{\updefault}$\Psi^{h_1}_1(w_1)$}}}}
\end{picture}%

%% file: overlap.pstex_t
\begin{picture}(0,0)%
\includegraphics{overlap.pstex}%
\end{picture}%
\setlength{\unitlength}{4144sp}%
\begingroup\makeatletter\ifx\SetFigFontNFSS\undefined%
\gdef\SetFigFontNFSS#1#2#3#4#5{%
  \reset@font\fontsize{#1}{#2pt}%
  \fontfamily{#3}\fontseries{#4}\fontshape{#5}%
  \selectfont}%
\fi\endgroup%
\begin{picture}(3060,2687)(1561,-6537)
\put(3852,-6405){\makebox(0,0)[lb]{\smash{{\SetFigFontNFSS{11}{13.2}{\familydefault}{\mddefault}{\updefault}$A$}}}}
\put(4445,-5765){\makebox(0,0)[lb]{\smash{{\SetFigFontNFSS{11}{13.2}{\familydefault}{\mddefault}{\updefault}$B$}}}}
\put(4559,-5085){\makebox(0,0)[lb]{\smash{{\SetFigFontNFSS{11}{13.2}{\familydefault}{\mddefault}{\updefault}$C$}}}}
\put(3692,-5425){\makebox(0,0)[lb]{\smash{{\SetFigFontNFSS{11}{13.2}{\familydefault}{\mddefault}{\updefault}$B'$}}}}
\put(3513,-5659){\makebox(0,0)[lb]{\smash{{\SetFigFontNFSS{11}{13.2}{\familydefault}{\mddefault}{\updefault}$A'$}}}}
\put(3779,-5152){\makebox(0,0)[lb]{\smash{{\SetFigFontNFSS{11}{13.2}{\familydefault}{\mddefault}{\updefault}$C'$}}}}
\put(2658,-6304){\makebox(0,0)[lb]{\smash{{\SetFigFontNFSS{11}{13.2}{\familydefault}{\mddefault}{\updefault}$.$}}}}
\put(2879,-6411){\makebox(0,0)[lb]{\smash{{\SetFigFontNFSS{11}{13.2}{\familydefault}{\mddefault}{\updefault}$.$}}}}
\put(3126,-6473){\makebox(0,0)[lb]{\smash{{\SetFigFontNFSS{11}{13.2}{\familydefault}{\mddefault}{\updefault}$.$}}}}
\put(3379,-6466){\makebox(0,0)[lb]{\smash{{\SetFigFontNFSS{11}{13.2}{\familydefault}{\mddefault}{\updefault}$.$}}}}
\put(2492,-6172){\makebox(0,0)[lb]{\smash{{\SetFigFontNFSS{11}{13.2}{\familydefault}{\mddefault}{\updefault}$.$}}}}
\put(3557,-4052){\makebox(0,0)[lb]{\smash{{\SetFigFontNFSS{11}{13.2}{\familydefault}{\mddefault}{\updefault}$.$}}}}
\put(3805,-4133){\makebox(0,0)[lb]{\smash{{\SetFigFontNFSS{11}{13.2}{\familydefault}{\mddefault}{\updefault}$.$}}}}
\put(4032,-4260){\makebox(0,0)[lb]{\smash{{\SetFigFontNFSS{11}{13.2}{\familydefault}{\mddefault}{\updefault}$.$}}}}
\put(4606,-5457){\makebox(0,0)[lb]{\smash{{\SetFigFontNFSS{11}{13.2}{\familydefault}{\mddefault}{\updefault}$\Psi^{h_k}_k(w_k)$}}}}
\put(4506,-4785){\makebox(0,0)[lb]{\smash{{\SetFigFontNFSS{11}{13.2}{\familydefault}{\mddefault}{\updefault}$\Psi^{h_{k+1}}_{k+1}(w_{k+1})$}}}}
\put(4345,-6099){\makebox(0,0)[lb]{\smash{{\SetFigFontNFSS{11}{13.2}{\familydefault}{\mddefault}{\updefault}$\Psi^{h_{k-1}}_{k-1}(w_{k-1})$}}}}
\put(1666,-4741){\makebox(0,0)[lb]{\smash{{\SetFigFontNFSS{11}{13.2}{\familydefault}{\mddefault}{\updefault}$\Psi^{h_1}_1(w_1)$}}}}
\put(3241,-5101){\makebox(0,0)[lb]{\smash{{\SetFigFontNFSS{11}{13.2}{\familydefault}{\mddefault}{\updefault}$\phi(0)$}}}}
\put(1576,-5506){\makebox(0,0)[lb]{\smash{{\SetFigFontNFSS{11}{13.2}{\familydefault}{\mddefault}{\updefault}$\Psi^{h_2}_2(w_2)$}}}}
\put(2746,-4021){\makebox(0,0)[lb]{\smash{{\SetFigFontNFSS{11}{13.2}{\familydefault}{\mddefault}{\updefault}$\Psi^{h_M}_M(w_M)$}}}}
\end{picture}%

%% file: 3wd.pstex_t
\begin{picture}(0,0)%
\includegraphics{3wd.pstex}%
\end{picture}%
\setlength{\unitlength}{4144sp}%
\begingroup\makeatletter\ifx\SetFigFontNFSS\undefined%
\gdef\SetFigFontNFSS#1#2#3#4#5{%
  \reset@font\fontsize{#1}{#2pt}%
  \fontfamily{#3}\fontseries{#4}\fontshape{#5}%
  \selectfont}%
\fi\endgroup%
\begin{picture}(4683,1099)(4531,-6209)
\put(8815,-5906){\makebox(0,0)[lb]{\smash{{\SetFigFontNFSS{11}{13.2}{\familydefault}{\mddefault}{\updefault}${\cal O}$}}}}
\put(8686,-5281){\makebox(0,0)[lb]{\smash{{\SetFigFontNFSS{11}{13.2}{\familydefault}{\mddefault}{\updefault}$\Psi_1$}}}}
\put(9181,-6136){\makebox(0,0)[lb]{\smash{{\SetFigFontNFSS{11}{13.2}{\familydefault}{\mddefault}{\updefault}$\Psi_3$}}}}
\put(8236,-6136){\makebox(0,0)[lb]{\smash{{\SetFigFontNFSS{11}{13.2}{\familydefault}{\mddefault}{\updefault}$\Psi_2$}}}}
\put(4927,-5675){\makebox(0,0)[lb]{\smash{{\SetFigFontNFSS{11}{13.2}{\familydefault}{\mddefault}{\updefault}${\cal O}$}}}}
\put(4996,-5281){\makebox(0,0)[lb]{\smash{{\SetFigFontNFSS{11}{13.2}{\familydefault}{\mddefault}{\updefault}$\Psi_1$}}}}
\put(4546,-6136){\makebox(0,0)[lb]{\smash{{\SetFigFontNFSS{11}{13.2}{\familydefault}{\mddefault}{\updefault}$\Psi_2$}}}}
\put(5491,-6136){\makebox(0,0)[lb]{\smash{{\SetFigFontNFSS{11}{13.2}{\familydefault}{\mddefault}{\updefault}$\Psi_3$}}}}
\put(6751,-5954){\makebox(0,0)[lb]{\smash{{\SetFigFontNFSS{11}{13.2}{\familydefault}{\mddefault}{\updefault}${\cal O}$}}}}
\put(6841,-5281){\makebox(0,0)[lb]{\smash{{\SetFigFontNFSS{11}{13.2}{\familydefault}{\mddefault}{\updefault}$\Psi_1$}}}}
\put(6391,-6136){\makebox(0,0)[lb]{\smash{{\SetFigFontNFSS{11}{13.2}{\familydefault}{\mddefault}{\updefault}$\Psi_2$}}}}
\put(7336,-6136){\makebox(0,0)[lb]{\smash{{\SetFigFontNFSS{11}{13.2}{\familydefault}{\mddefault}{\updefault}$\Psi_3$}}}}
\put(5896,-5821){\makebox(0,0)[lb]{\smash{{\SetFigFontNFSS{11}{13.2}{\familydefault}{\mddefault}{\updefault}$+$}}}}
\put(7786,-5821){\makebox(0,0)[lb]{\smash{{\SetFigFontNFSS{11}{13.2}{\familydefault}{\mddefault}{\updefault}$+$}}}}
\end{picture}%

%% file: OB.pstex_t
\begin{picture}(0,0)%
\includegraphics{OB.pstex}%
\end{picture}%
\setlength{\unitlength}{4144sp}%
\begingroup\makeatletter\ifx\SetFigFontNFSS\undefined%
\gdef\SetFigFontNFSS#1#2#3#4#5{%
  \reset@font\fontsize{#1}{#2pt}%
  \fontfamily{#3}\fontseries{#4}\fontshape{#5}%
  \selectfont}%
\fi\endgroup%
\begin{picture}(5581,881)(5023,-6199)
\put(5287,-5675){\makebox(0,0)[lb]{\smash{{\SetFigFontNFSS{11}{13.2}{\familydefault}{\mddefault}{\updefault}${\cal O}$}}}}
\put(8455,-5906){\makebox(0,0)[lb]{\smash{{\SetFigFontNFSS{11}{13.2}{\familydefault}{\mddefault}{\updefault}${\cal O}$}}}}
\put(6751,-5954){\makebox(0,0)[lb]{\smash{{\SetFigFontNFSS{11}{13.2}{\familydefault}{\mddefault}{\updefault}${\cal O}$}}}}
\put(6121,-5821){\makebox(0,0)[lb]{\smash{{\SetFigFontNFSS{11}{13.2}{\familydefault}{\mddefault}{\updefault}$+$}}}}
\put(7606,-5821){\makebox(0,0)[lb]{\smash{{\SetFigFontNFSS{11}{13.2}{\familydefault}{\mddefault}{\updefault}$+$}}}}
\put(9226,-5821){\makebox(0,0)[lb]{\smash{{\SetFigFontNFSS{11}{13.2}{\familydefault}{\mddefault}{\updefault}$=$}}}}
\put(9991,-5641){\makebox(0,0)[lb]{\smash{{\SetFigFontNFSS{11}{13.2}{\familydefault}{\mddefault}{\updefault}${\cal O}$}}}}
\end{picture}%

%% file: stripsC3.pstex_t
\begin{picture}(0,0)%
\includegraphics{stripsC3.pstex}%
\end{picture}%
\setlength{\unitlength}{4144sp}%
\begingroup\makeatletter\ifx\SetFigFontNFSS\undefined%
\gdef\SetFigFontNFSS#1#2#3#4#5{%
  \reset@font\fontsize{#1}{#2pt}%
  \fontfamily{#3}\fontseries{#4}\fontshape{#5}%
  \selectfont}%
\fi\endgroup%
\begin{picture}(5675,2572)(3338,-7661)
\put(4562,-5803){\makebox(0,0)[lb]{\smash{{\SetFigFontNFSS{11}{13.2}{\familydefault}{\mddefault}{\updefault}$=$}}}}
\put(5920,-5803){\makebox(0,0)[lb]{\smash{{\SetFigFontNFSS{11}{13.2}{\familydefault}{\mddefault}{\updefault}$+$}}}}
\put(7718,-5815){\makebox(0,0)[lb]{\smash{{\SetFigFontNFSS{11}{13.2}{\familydefault}{\mddefault}{\updefault}$-$}}}}
\put(3781,-5851){\makebox(0,0)[lb]{\smash{{\SetFigFontNFSS{20}{24.0}{\familydefault}{\mddefault}{\updefault}$V^\epsilon_3$}}}}
\put(7516,-7306){\makebox(0,0)[lb]{\smash{{\SetFigFontNFSS{20}{24.0}{\familydefault}{\mddefault}{\updefault}$C^\epsilon_3$}}}}
\put(6931,-5866){\makebox(0,0)[lb]{\smash{{\SetFigFontNFSS{20}{24.0}{\familydefault}{\mddefault}{\updefault}$V^\epsilon_3$}}}}
\end{picture}%

%% file: stripsC4.pstex_t
\begin{picture}(0,0)%
\includegraphics{stripsC4.pstex}%
\end{picture}%
\setlength{\unitlength}{4144sp}%
\begingroup\makeatletter\ifx\SetFigFontNFSS\undefined%
\gdef\SetFigFontNFSS#1#2#3#4#5{%
  \reset@font\fontsize{#1}{#2pt}%
  \fontfamily{#3}\fontseries{#4}\fontshape{#5}%
  \selectfont}%
\fi\endgroup%
\begin{picture}(5301,1243)(7119,-7870)
\put(10122,-7273){\makebox(0,0)[lb]{\smash{{\SetFigFontNFSS{11}{13.2}{\familydefault}{\mddefault}{\updefault}$t$}}}}
\put(11770,-7806){\makebox(0,0)[lb]{\smash{{\SetFigFontNFSS{11}{13.2}{\familydefault}{\mddefault}{\updefault}$t$}}}}
\put(10523,-7252){\makebox(0,0)[lb]{\smash{{\SetFigFontNFSS{11}{13.2}{\familydefault}{\mddefault}{\updefault}$\displaystyle{+ \, \int_0^{2 \epsilon}dt}$}}}}
\put(7516,-7306){\makebox(0,0)[lb]{\smash{{\SetFigFontNFSS{20}{24.0}{\familydefault}{\mddefault}{\updefault}$C^\epsilon_4$}}}}
\put(8520,-7257){\makebox(0,0)[lb]{\smash{{\SetFigFontNFSS{11}{13.2}{\familydefault}{\mddefault}{\updefault}$\displaystyle{= \, \int_{2 \epsilon}^0dt}$}}}}
\end{picture}%

%% file: loops.pstex_t
\begin{picture}(0,0)%
\includegraphics{loops.pstex}%
\end{picture}%
\setlength{\unitlength}{4144sp}%
\begingroup\makeatletter\ifx\SetFigFont\undefined%
\gdef\SetFigFont#1#2#3#4#5{%
  \reset@font\fontsize{#1}{#2pt}%
  \fontfamily{#3}\fontseries{#4}\fontshape{#5}%
  \selectfont}%
\fi\endgroup%
\begin{picture}(2848,2136)(3519,-5151)
\put(5121,-3162){\makebox(0,0)[lb]{\smash{{\SetFigFont{11}{13.2}{\familydefault}{\mddefault}{\updefault}$g_2$}}}}
\put(5361,-3936){\makebox(0,0)[lb]{\smash{{\SetFigFont{11}{13.2}{\familydefault}{\mddefault}{\updefault}$t_1$}}}}
\put(5007,-3736){\makebox(0,0)[lb]{\smash{{\SetFigFont{11}{13.2}{\familydefault}{\mddefault}{\updefault}$t_2$}}}}
\put(4220,-4116){\makebox(0,0)[lb]{\smash{{\SetFigFont{11}{13.2}{\familydefault}{\mddefault}{\updefault}$t_i$}}}}
\put(5374,-4716){\makebox(0,0)[lb]{\smash{{\SetFigFont{11}{13.2}{\familydefault}{\mddefault}{\updefault}$t_n$}}}}
\put(5874,-3489){\makebox(0,0)[lb]{\smash{{\SetFigFont{11}{13.2}{\familydefault}{\mddefault}{\updefault}$g_1$}}}}
\put(3534,-3869){\makebox(0,0)[lb]{\smash{{\SetFigFont{11}{13.2}{\familydefault}{\mddefault}{\updefault}$g_i$}}}}
\put(5975,-4989){\makebox(0,0)[lb]{\smash{{\SetFigFont{11}{13.2}{\familydefault}{\mddefault}{\updefault}$g_n$}}}}
\put(3601,-5096){\makebox(0,0)[lb]{\smash{{\SetFigFont{11}{13.2}{\familydefault}{\mddefault}{\updefault}$h$}}}}
\end{picture}%